\documentclass[useAMS,usenatbib,twocolumn]{mnras}

\usepackage[usenames,dvipsnames]{xcolor}

\usepackage{amssymb, amsmath, mathtools}
\usepackage{epsf}
\usepackage{bm}
\usepackage{ulem}

\usepackage{natbib}
\usepackage{graphicx}
\usepackage{cancel}

\def\la{\; \raise0.3ex\hbox{$<$\kern-0.75em\raise-1.1ex\hbox{$\sim$}}\;}
\def\ga{\;  \raise0.3ex\hbox{$>$\kern-0.75em\raise-1.1ex\hbox{$\sim$}}\;}

\newcommand*{\ci}[1]{_{\mkern-3.5mu#1}}

\title[Force on proton vortices in neutron stars]
{Force on proton vortices in superfluid neutron stars}
\author[M. E. Gusakov]
{M. E. Gusakov\thanks{gusakov@astro.ioffe.ru}
\\
Ioffe Institute,
Polytekhnicheskaya 26, 194021 St.-Petersburg, Russia
}

\begin{document}

\date{Accepted 2019 xxxx. Received 2019 xxxx; in original form 2019 xxxx}

\pagerange{\pageref{firstpage}--\pageref{lastpage}} \pubyear{2019}

\maketitle

\label{firstpage}

%
\begin{abstract}
Force on proton vortices in superfluid and superconducting 
matter of neutron stars  
is calculated 
at
vanishing stellar temperature.
Both longitudinal (dissipative) and transverse (Lorentz-type) components of the force
are derived in a coherent way and compared in detail with the 
corresponding expressions
available in the literature.
This allows us to resolve a controversy 
about the form of the Lorentz-type force component
acting on proton vortices.
The calculated force 
is a key ingredient in magnetohydrodynamics of superconducting neutron stars
and is important for modeling the evolution of stellar magnetic field.
\end{abstract}
%


\maketitle

\section{Introduction and formulation of the problem}
\label{intro}

The magnetic field in neutron stars (NSs) varies in a very wide range 
from $\sim (10^8 -10^{9})$~G in millisecond pulsars and neutron stars in low-mass X-ray binaries
to $\sim 10^{12}$~G in ordinary radio pulsars and up to $\sim 10^{15}$~G in magnetars \citep{kaspi10,vrppam13}.
It is a challenge for theorists to explain such diverse objects within 
a unified theoretical model.  
The problem is significantly complicated by the fact that
NS matter can become superfluid/superconducting at stellar temperatures $T \sim (10^8-10^{10})$~K \citep{plps13,gps14,sc18}.
The magnetic field in such matter is confined to proton vortices, 
also called Abrikosov vortices or proton flux tubes (\citealt*{bpp69a,sauls89}).%
%
\footnote{We assume that protons form a type-II superconductor, 
see Section \ref{vort} for more details.}
%
Therefore, to describe the evolution of the magnetic field in NSs
it is necessary to understand in detail 
the vortex dynamics,
a very complicated problem, full of controversies in the literature,
which has not been fully solved yet (see \citealt{hs17, sc18} for recent reviews). 

Here we would like to focus on one such controversy 
related to the forces acting on proton vortices in superconducting  NSs.
In what follows, to make our analysis as simple as possible, 
we 
consider a strongly degenerate npe-matter of NS cores composed of superfluid neutrons (n),
superconducting protons (p), and electrons (e) 
(the effect of muons will be discussed in Section \ref{disc}).
For simplicity, we assume that the temperature $T$ is so small that there
are almost no thermal Bogoliubov neutron and proton excitations in the system in the absence of vortices.
We also neglect the effects of neutron-proton entrainment,
assuming that the off-diagonal elements of the entrainment matrix $\rho_{ik}$ 
vanish, $\rho_{\rm np}=\rho_{\rm pn}=0$ (see, e.g., \citealt{ab76} for a definition of $\rho_{ik}$). 
In principle, all these simplifying assumptions, 
except for the assumption $T=0$, 
can be easily relaxed.
However, extension of our results to finite temperatures 
is more intricate, since it requires a detailed understanding 
of how vortices interact with neutron and proton thermal Bogoliubov excitations
-- an almost unexplored problem in the context of NS physics 
(see \citealt{kopnin02, sonin16} for a general approach to attack it).

\subsection*{Description of the controversy  }

We shall start with the equation, describing the magnetic field evolution
in superconducting NSs \citep*{kg00,gd16,dg17,blb17},
\begin{align}
&
\frac{\partial \overline{{\pmb B}}}{\partial t}
={\pmb \nabla}\times ({\pmb V}_{\rm L}\times \overline{{\pmb B}}),
&
\label{Bevol}
\end{align}
where $\overline{{\pmb B}}$ is the stellar magnetic field, 
averaged over the volume containing many vortices
(more precisely, it is the magnetic induction field);
${\pmb V}_{\rm L}$ is the local vortex velocity.
This equation simply states, that the magnetic field, confined to proton vortices,
is transported with the vortex velocity, ${\pmb V}_{\rm L}$.
To solve (\ref{Bevol}) one needs first to define ${\pmb V}_{\rm L}$,
i.e., to express it
through available transport particle velocities in the system.
For example, for 
npe-mixture at zero temperature ($T=0$),
the relevant velocities are the electron velocity, ${\pmb u}_{\rm e}$,
and the superfluid neutron and proton velocities, ${\pmb V}_{\rm sn}$ and ${\pmb V}_{\rm sp}$, respectively
(see Section \ref{formulation} for an accurate definition of these velocities).
To express ${\pmb V}_{\rm L}$ through ${\pmb u}_{\rm e}$, ${\pmb V}_{\rm sn}$, and ${\pmb V}_{\rm sp}$,
one should write down a force balance equation for a vortex,
which can be customarily presented in the form (e.g., \citealt*{gas11}):
\begin{align}
&
{\pmb F}_{\rm buoyancy}+{\pmb F}_{\rm tension}
+{\pmb F}_{\rm npe \rightarrow V}=0.
&
\label{forcebalance0}
\end{align}
In writing this equation we assumed that the mass of the vortex per unit length 
is negligible, so that the sum of the forces (per unit length) 
on a vortex must vanish \citep{donnelly05}.
In equation (\ref{forcebalance0})  
${\pmb F}_{\rm buoyancy}$ and ${\pmb F}_{\rm tension}$ are the buoyancy and tension forces, respectively
(their actual form is not important for us here; see, e.g., \citealt{dg17} for details;
in Section \ref{formulation} these forces, 
which do not depend on transport velocities, 
will be denoted ${\pmb F}_{\rm ext}$);
${\pmb F}_{\rm npe \rightarrow V}$ is the {\it total} velocity-dependent force on a vortex from npe-matter.
Taking into account the so called `screening condition', 
${\pmb V}_{\rm sp}={\pmb u}_{\rm e}$, 
which should be satisfied in NS bulk,
and neglecting entrainment effects,
${\pmb F}_{\rm npe\rightarrow V}$ takes the form
(see Section \ref{formulation} for a detailed derivation),
\begin{align}
&
{\pmb F}_{\rm npe\rightarrow V} = 
-D [{\pmb e}_z \times [{\pmb e}_z\times({\pmb u}_e-{\pmb V}_{\rm L})]]
+D'[{\pmb e}_z\times ({\pmb u}_{\rm e}-{\pmb V}_{\rm L})].
&
\label{FnpeV}
\end{align}
The first term here
describes longitudinal (dissipative) 
part of the force, 
the second term 
is 
the transverse 
(Lorentz-like)
part. 
In equation (\ref{FnpeV})
${\pmb e}_{z}$ is the unit vector 
along the vortex line (see Section \ref{vort});
the kinetic coefficients $D$ and $D'$ should be determined from the microscopic theory.

At this point we face a controversy in the literature regarding 
the value of the coefficient $D'$ in equation (\ref{FnpeV}).%
%
\footnote{It turns out 
	that there are also no agreement about the value 
	of the coefficient $D$ 
	in the literature (see Section \ref{disc}).
}
%
According to \cite{jones91,jones06} $D'=0$.
His result 
is based on the following arguments. 
Superconducting protons act on a vortex with the Magnus force 
(e.g., \citealt{nv66,kopnin02};
\citealt{gas11}),
\begin{align}
&
{\pmb F}_{\rm M} = -\pi \hbar\,  n_{\rm p} \, \left[
{\pmb e}_{z}\times ({\pmb V}_{\rm sp}-{\pmb V}_{\rm L})
\right] 
&
\nonumber\\
&\quad \,\,\,= -\pi \hbar\,  n_{\rm p} \, \left[
{\pmb e}_{z}\times ({\pmb u}_{\rm e}-{\pmb V}_{\rm L})
\right],&
\label{Fmagnus}
\end{align}
where $n_{\rm p}$ is the proton number density
and in the second equality we made use of the screening condition, 
${\pmb V}_{\rm sp}={\pmb u}_{\rm e}$.
Note that
the Magnus force (\ref{Fmagnus})
coincides with the Lorentz force on protons, 
${\pmb F}_{\rm Lp} =(1/c)\,  [{\pmb J}_{\rm p}\times {\pmb \Phi_0}]$,
where ${\pmb J}_{\rm p}=e_{\rm p} n_{\rm p}({\pmb V}_{\rm sp}-{\pmb V}_{\rm L})$
is the proton current density in the coordinate system in which ${\pmb V}_{\rm L}=0$;
$e_{\rm p}$ is the proton charge;
${\pmb \Phi_0}=\Phi_0 \, {\pmb e}_{ z}$
is the vector directed along ${\pmb e}_z$, 
whose absolute value 
equals
the 
total magnetic flux of a proton vortex, $\Phi_0=\pi \hbar c/e_{\rm p} 
\approx 2.07 \times 10^{-7}$~G~cm$^2$.
The fact that protons act on a vortex 
with the Lorentz force ${\pmb F}_{\rm Lp}(={\pmb F}_{\rm M}$)
may 
lead to idea
that electrons also act on a vortex with the corresponding Lorentz force, 
${\pmb F}_{\rm Le} =(1/c) \, [{\pmb J}_{\rm e}\times {\pmb \Phi_0}]$,
where ${\pmb J}_{\rm e}=e_{\rm e} n_{\rm e}({\pmb u}_{\rm e}-{\pmb V}_{\rm L})$, 
and
$e_{\rm e}$, $n_{\rm e}$ are the electron charge and number densities, respectively.
Sum of these two forces, ${\pmb F}_{\rm Lp}+{\pmb F}_{\rm Le}$, equals zero 
(\citealt{jones09})
because 
of the screening and quasineutrality conditions ($n_{\rm e}=n_{\rm p}$),
which allows Jones to conclude 
that the total transverse force on a vortex vanishes, $D'=0$.

Unlike Jones, \cite{as10} postulated (without justification)
that $D'=-\pi \hbar n_{\rm p}$, 
i.e., the only transverse force on a vortex is the Magnus force, 
${\pmb F}_{\rm M}$.
In turn, \cite{gas11} also assumed that $D'=-\pi \hbar n_{\rm p}$,
arguing that there are three transverse forces acting on a vortex in npe-matter,
namely, 
the Magnus force ${\pmb F}_{\rm M}$, 
the (minus) electron Lorentz force, $-{\pmb F}_{\rm Le}$, 
and (minus) proton Lorentz force, $-{\pmb F}_{\rm Lp}$ 
(the minus sign appears due to the Newton's third law; for example, 
electrons are subject to the force ${\pmb F}_{\rm Le}$ 
in the magnetic field of a vortex, 
thus the force on a vortex is $-{\pmb F}_{\rm Le}$).
Sum of these forces, ${\pmb F}_{\rm M}+(-{\pmb F}_{\rm Le})+(-{\pmb F}_{\rm Lp})$
equals ${\pmb F}_{\rm M}$, hence $D'=-\pi \hbar n_p$.

Both interpretations of \cite{jones91,jones06} and \cite{gas11} are not very convincing.
The interpretation by Jones have an obvious problem with the Newton's third law:
Electrons act on a vortex with the Lorentz force ${\pmb F}_{\rm Le}$ on electrons, which is 
strange.
The interpretation by Glampedakis et al.\ is also 
confusing,
because 
it assumes that the Magnus and Lorentz forces ${\pmb F}_{\rm M}$ and ${\pmb F}_{\rm Lp}$ 
are of different origin, although it is, in fact, 
two different names for the same force (\citealt{nv66}).

So, what is the correct value of $D'$?
The answer is very important since the vortex velocity ${\pmb V}_{\rm L}$,
defined by equation (\ref{forcebalance0}),
can vary by orders of magnitude depending on the choice of $D'$.
This uncertainty can affect dramatically the 
typical magnetic field evolution timescales 
(see equation \ref{Bevol} and compare
the evolution timescales, e.g., in \citealt{jones06};~\citealt{blb17} 
and in \citealt{gagl15};~\citealt{eprgv15}).

The present work is devoted to answering this question.
In Section \ref{vort} we discuss the basic parameters characterizing proton vortices
and lengthscales that play a role in our problem.
In Section \ref{formulation} we derive a basic expression for the total force on a vortex.
Instead of considering separate contributions to the force from each particle species
(a way, which apparently leads to contradictory results in the literature), we decided to 
extract the force on a vortex from the analysis of total momentum conservation equation 
for the system as a whole. 
This derivation method 
is inspired by 
the work of \cite{sonin76,gs76,aggk81}.
Further, in Section \ref{cross_section} we calculate two necessary cross-sections,
which determine the coefficients $D$ and $D'$.
We discuss the obtained force on a vortex and compare it with the results
available in the literature in Section \ref{disc}.
Finally, we conclude in Section \ref{concl}.

\section{Basic parameters and hierarchy of lengthscales}
\label{vort}

\begin{figure}
	\begin{center}
		\includegraphics[width=0.3\textwidth]{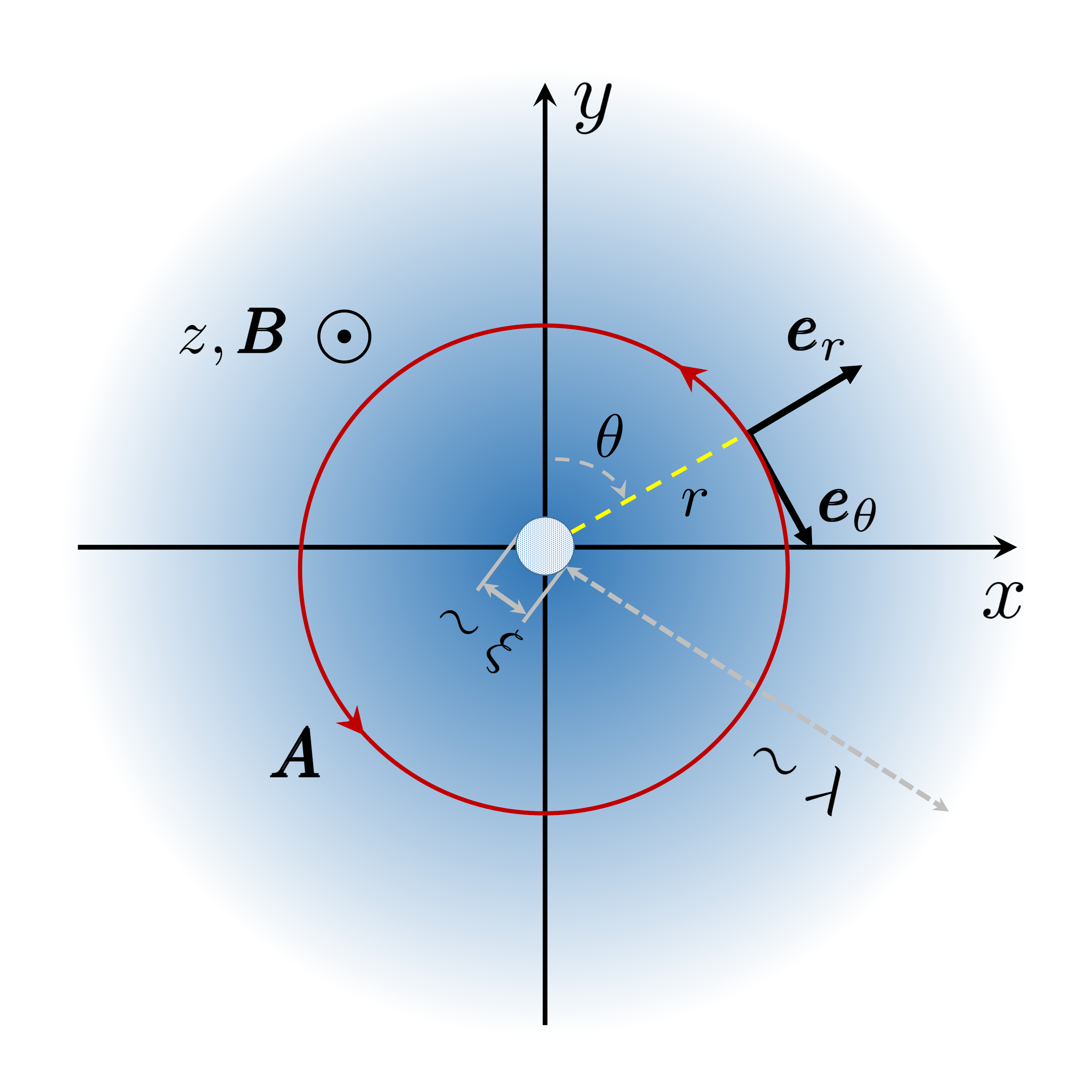}
		\caption{\label{scheme}
			Scheme of a proton vortex. The vortex magnetic field is directed along the axis $z$.
			Centre of the vortex corresponds to $x=y=0$.
		}
	\end{center}
\end{figure}

Schematically, the proton vortex consists of the `normal' core%
%
\footnote{Inside the vortex core proton quasiparticles exist even at $T=0$ \citep*{cdm64}.}
%
with the radius of the order of the coherence length, $\xi$,
surrounded by the more extended region containing the magnetic field (see Fig.~\ref{scheme}). 
The radius of that region is $\sim \lambda$, 
where $\lambda$ is the London penetration depth.
The parameters $\xi$ and $\lambda$ are given by the formulas (e.g., \citealt{ll80,degennes99})
\begin{align}
&
\xi = \frac{\hbar p_{\rm Fp}}{\pi m_{\rm p}^\ast \Delta_{\rm p}}
\approx 28  \, {\rm fm} \left(\frac{n_{\rm p}}{0.18 n_0} \right)^{1/3}
\left(\frac{m_{\rm p}}{m_{\rm p}^\ast}\right) \left(\frac{0.456 \, {\rm MeV}}{\Delta_{\rm p}}\right),
&
\label{xi}\\
&
\lambda = \sqrt{\frac{m_{\rm p} c^2}{4 \pi e_{\rm p}^2 n_{\rm sp}}}
\approx 42.4 \, {\rm fm} 
\left(\frac{0.18 n_0}{n_{\rm sp}} \right)^{1/2}.
&
\label{lambda}
\end{align}
Here 
$p_{\rm Fp}$, 
$m_{\rm p}$, $m_{\rm p}^\ast$ 
are the proton Fermi momentum, 
mass, and effective mass, respectively;
$n_0=0.16$~fm$^{-3}$ is the nuclear matter density;
$n_{\rm sp}(T)$ and $\Delta_{\rm p}(T)$ are the superfluid proton number density and energy gap, respectively.
At $T=0$ one has $n_{\rm sp}=n_{\rm p}$ and $\Delta_{\rm p} \approx k_{\rm B} T_{\rm cp}/0.567$, 
where $T_{\rm cp}$ is the proton critical temperature. 
In particular, 
$\Delta_{\rm p} \approx 0.456$~MeV
for 
$T_{\rm cp}=3 \times 10^9$~K. 

Note that, generally, 
$\xi(T)$ can be comparable to $\lambda(T)$ for NS conditions,
but
proton vortices may exist only for type-II superconductors, 
for which $\xi(T)<\sqrt{2}\lambda(T)$ \citep{ll80,degennes99}.
This condition can be violated in the deep layers of NS cores (e.g., \citealt{sedrakyan05,jones06,gas11,gd16}).

%
%
Following \cite*{als84}, we parametrize the vortex magnetic field ${\pmb B}(r)$ as
\begin{align}
&
{\pmb B}={\pmb e}_z  \left(\frac{\Phi}{\pi \xi^2}\right) \cdot
\begin{dcases}
1-\frac{\xi}{\lambda} \, K_1\left(\frac{\xi}{\lambda}\right) I_0\left(\frac{r}{\lambda}\right),& \,\,\, 0\leq r <\xi; \\
\frac{\xi}{\lambda} \, I_1\left(\frac{\xi}{\lambda}\right) K_0\left(\frac{r}{\lambda}\right), & \,\,\, r\geq \xi,
\end{dcases}
&
\label{B}
\end{align}
where $\Phi$ is the magnetic flux associated with the vortex line; for a proton vortex 
$\Phi=\Phi_0$ (see Section \ref{intro} for a definition of $\Phi_0$;
we emphasize that the results obtained in this paper 
are presented in the form valid for arbitrary magnetic flux of a vortex).
The corresponding vector potential ${\pmb A}(r)$ is 
(we work in the cylindrical coordinate system $(r,\,\theta,\, z)$,
defined in Fig.\ \ref{scheme})%
\begin{align}
&
{\pmb A}=-{\pmb e}_\theta  \left(\frac{\Phi}{2 \pi r}\right) \cdot
\begin{dcases}
\frac{r}{\xi}\left[\frac{r}{\xi}-2 K_1\left(\frac{\xi}{\lambda}\right)I_1\left(\frac{r}{\lambda}\right)\right],& \,\, 0\leq r <\xi; \\
1-\frac{2r}{\xi}I_1\left(\frac{\xi}{\lambda}\right)K_1\left(\frac{r}{\lambda}\right), & \,\, r\geq \xi.
\end{dcases}
&
\label{Avec}
\end{align}
%
\begin{figure}
	\begin{center}
		\includegraphics[width=0.5\textwidth]{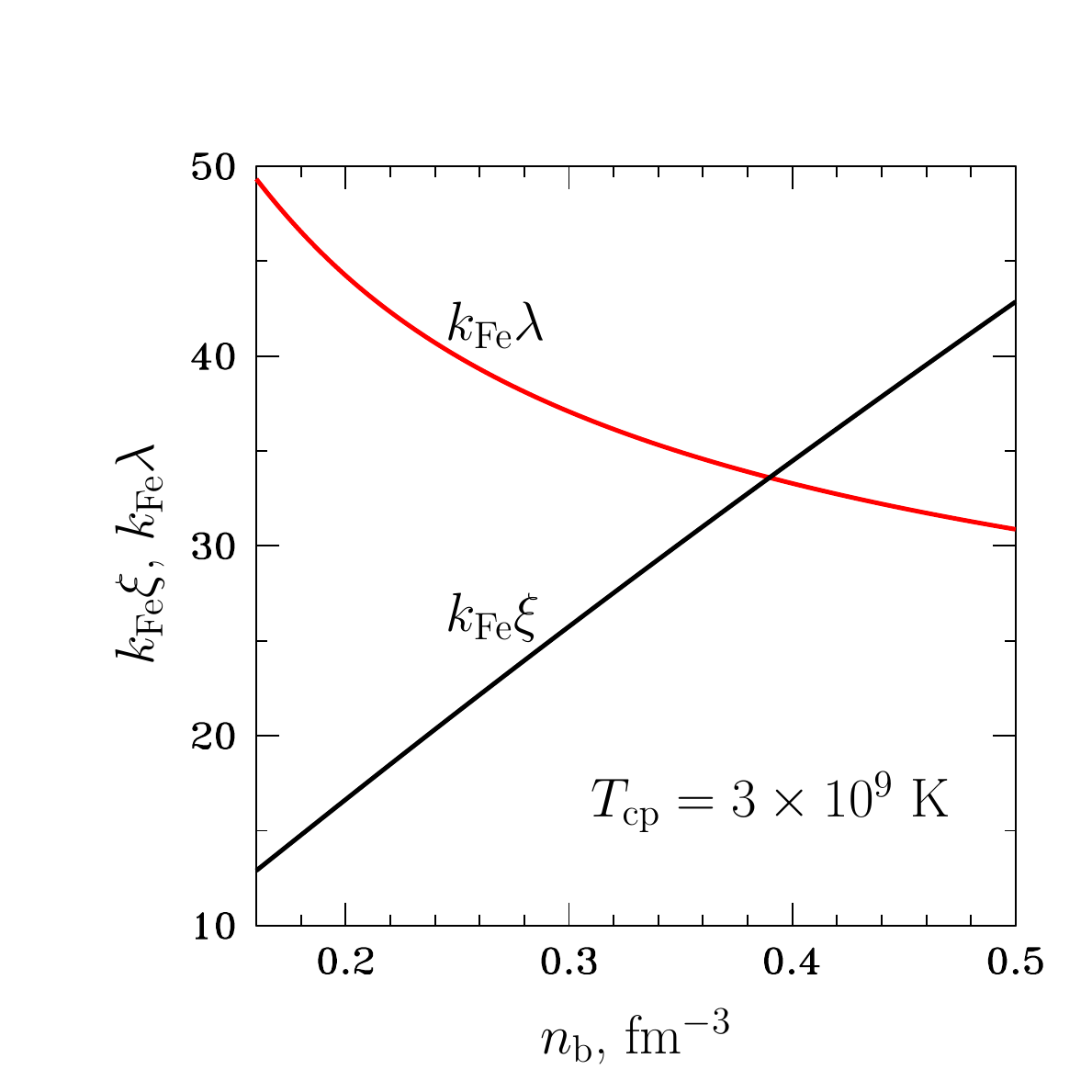}
		\caption{\label{param}
			Dimensionless parameters $k_{\rm Fe}\xi$ and $k_{\rm Fe}\lambda$ 
			versus baryon number density $n_{\rm b}$ for $T_{\rm cp}=3 \times 10^9$~K and $T=0$.
		}
	\end{center}
\end{figure}
%
Further it will be convenient to rewrite it as 
\begin{align}
&
{\pmb A} \equiv A_{\theta}\, {\pmb e}_\theta= -\frac{\hbar c}{e_{\rm p}} \, \frac{\zeta}{r} \, \mathcal{P}(r) \, {\pmb e}_\theta,
&
\label{A}
\end{align}
where
\begin{align}
&
\zeta=\frac{e_{\rm p} \Phi}{2 \pi \hbar c} = \frac{\Phi}{2 \Phi_0}
&
\label{gamma1}
\end{align}
and the function $\mathcal{P}(r)$ is defined by equation (\ref{Avec})
and is related to the magnetic field ${\pmb B}={\pmb \nabla}\times {\pmb A}$ 
by the following obvious equation: 
\begin{align}
&
{\pmb B}={\pmb e}_z \, \frac{\hbar c}{e_{\rm p}} \, 
\frac{\zeta}{r} \, \frac{{\rm d}\mathcal{P}(r)}{{\rm d}r}.
&
\label{BBB}
\end{align}
In equations (\ref{B})--(\ref{A}) 
${\pmb e}_\theta$ is the unit vector shown in Fig.\ \ref{scheme} and ${\pmb e}_{z}={\pmb B}/B$
is the unit vector in the direction of the magnetic field.
The important characteristic of magnetized NSs
is the average distance between neighbouring vortices, $d_{\rm B}$ (e.g., \citealt{tinkham96,degennes99})
\begin{align}
&
d_{\rm B}=\sqrt{\frac{2\Phi}{\sqrt{3} \, \overline{B}}}
=4.89\times 10^{3}~{\rm fm} \, \sqrt{\frac{10^{12} \, {\rm G}}{\overline{B}}},
&
\label{dB}
\end{align}
which is defined by specifying the average stellar magnetic induction field $\overline{B}$.
In this work we assume that $d_{\rm B}$ is larger than $\lambda$, i.e., 
we consider NSs with $\overline{B}$ smaller than
\begin{align}
&
\overline{B}<\frac{8 \pi \Phi e_{\rm p}^2 n_{\rm sp}}{\sqrt{3}m_{\rm p}c^2}
\approx 1.33\times 10^{16}\, {\rm G} 
\,\left(\frac{n_{\rm sp}}{0.18 n_0}\right)\left(\frac{\Phi}{\Phi_0}\right).
&
\label{Bcr}
\end{align}
Another important parameter in our problem is the typical electron wavelength (divided by $2\pi$)
\begin{align}
&
\frac{1}{k_{\rm Fe}}= 1.05 \,{\rm fm} 
\left(\frac{0.18 n_0}{n_{\rm e}}\right)^{1/3},
&
\label{wavelength}
\end{align}
where $k_{\rm Fe}=p_{\rm Fe}/\hbar$ with $p_{\rm Fe}$ 
being the electron Fermi momentum.
One sees that it is much smaller than  $\xi$, $\lambda$, and $d_{\rm B}$.
This enables us to study the forces acting on an isolated vortex
within the quasiclassical approximation and  
ignoring the presence of other vortices, i.e., the collective effects.%
%
\footnote{Note that the scattering cross-sections, calculated in Section \ref{cross_section},
which have a dimension of length, are also much smaller than $d_{\rm B}$.}
%
The dimensionless parameters $k_{\rm Fe} \xi$ and $k_{\rm Fe}\lambda$ 
are plotted in Fig.\ \ref{param}
as functions of the baryon number density $n_{\rm b}$. 
Here and below to plot the figures we employ the equation of state HHJ (\citealt{hhj99}) 
and assume $T_{\rm cp}=3 \times 10^9$~K.

Finally, the last important parameter that should be mentioned here
is the electron mean free path, $l$.
Generally, it is much larger than $d_{\rm B}$ (e.g., \citealt{ss18}) 
and hence than other typical lengthscales
discussed above even for normal (nonsuperfluid and nonsuperconducting) matter.
Nucleon superfluidity further increases $l$.
This means that at distances $r\ll l$ a perturbation of the electron distribution function
caused by the vortex can be found from the {\it collisionless} kinetic equation for electrons.
This property will be used in the next section. 

Summarizing, there are five relevant lengthscales, 
$\xi$, $\lambda$, $d_{\rm B}$, $1/k_{\rm Fe}$, and $l$, 
in the problem of calculation of the force acting on a vortex, 
and for typical NS conditions they are related by the inequality
\begin{align}
&
l \gg d_{\rm B} \gg \lambda \gtrsim \xi \gg 1/k_{\rm Fe}.
&
\label{ineq}
\end{align}	
%

\section{General expression for the force on a proton vortex }
\label{formulation}

\begin{figure}
	\begin{center}
		\includegraphics[width=0.3\textwidth]{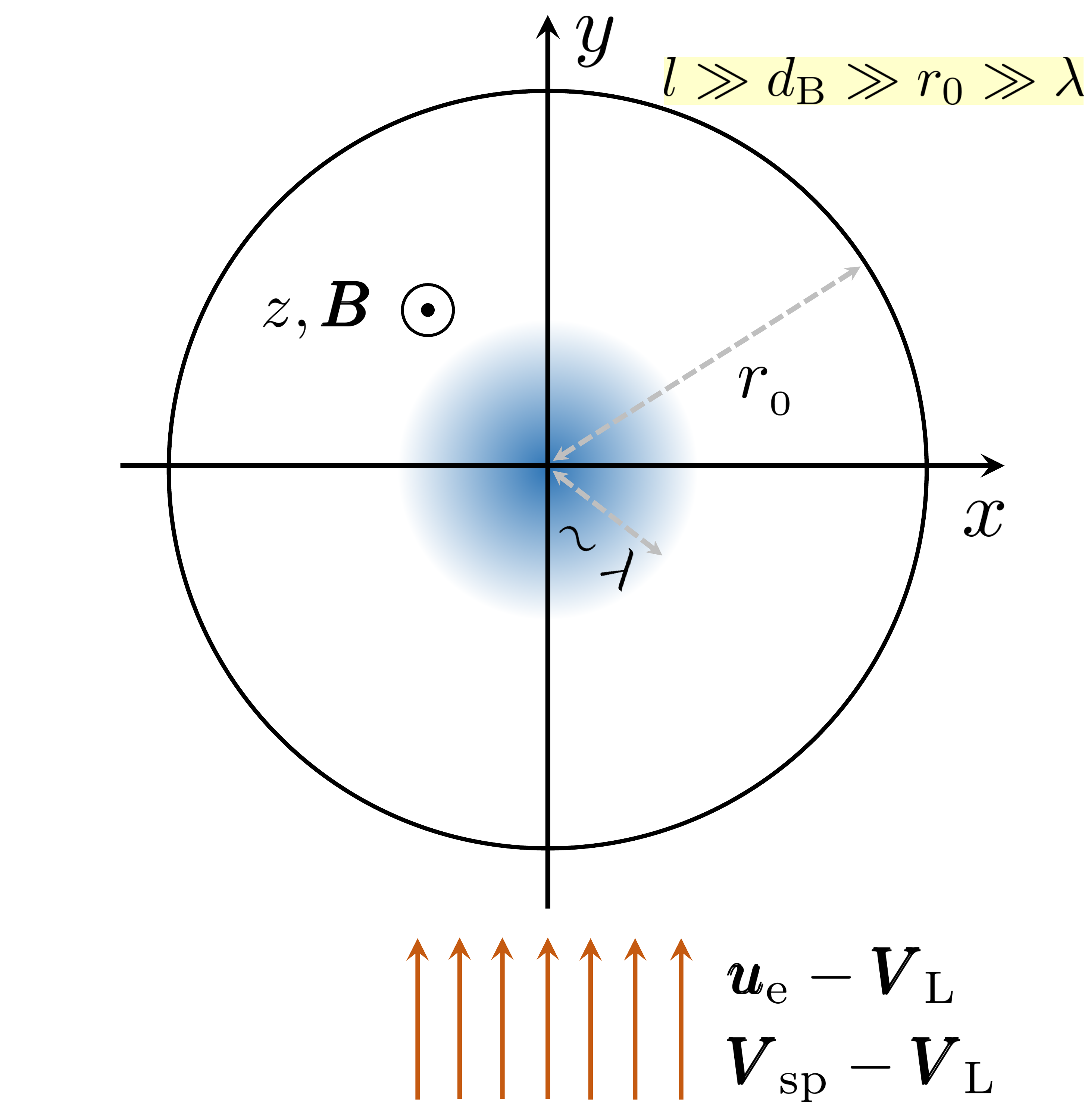}
		\caption{\label{vortvel}
			Top view on the vortex and a cylinder of radius $r_0$ used to take the integral (\ref{cond1}).
		}
	\end{center}
\end{figure}

Let us create a straight proton vortex (also called the Abrikosov vortex or flux tube)
in the initially homogeneous system, 
with the magnetic field directed along the axis $z$, as shown in Fig.~\ref{scheme}. 
As in Section \ref{intro}, 
the vortex velocity is ${\pmb V}_{\rm L}$;
the velocities of neutrons, protons, and electrons 
far from the vortex are denoted as 
${\pmb V}_{{\rm sn}}$, ${\pmb V}_{{\rm sp}}$, 
and ${\pmb u}_{\rm e}$, respectively.
Because of the screening condition (\citealt{jones91,jones06,gas11,gd16})
the electron and proton currents (and hence velocities) must 
coincide
to a very high precision in the bulk of NS superconductor,
\begin{align}
&{\pmb V}_{{\rm sp}}={\pmb u}_{\rm e}.&
\label{screening}
\end{align}
Our aim is to calculate the velocity-dependent force (per unit length), 
which acts on the vortex from the surrounding matter.
Below in Sections \ref{formulation} and \ref{cross_section}
we assume that electrons scatter only on the magnetic field of a vortex
and do not scatter on the localized proton excitations in the vortex core.
The effect of the latter type of scattering will be discussed in Section \ref{disc}.
Taking into account the condition (\ref{screening}) the force can be, quite generally, 
written as (e.g., \citealt{donnelly05, sonin16})
\begin{align}
&
{\pmb F}_{{\rm npe}\rightarrow {\rm V}}=
-D [{\pmb e}_z \times [{\pmb e}_z\times({\pmb u}_e-{\pmb V}_{\rm L})]]
+D'[{\pmb e}_z\times ({\pmb u}_{\rm e}-{\pmb V}_{\rm L})]
&
\nonumber\\
&
\quad\quad\quad\,\,\,+D_z \, {\pmb e}_z \, [{\pmb e}_z \cdot ({\pmb u}_{\rm e}-{\pmb V}_{\rm L})],
&
\label{force}
\end{align}
where we accounted for the fact that 
the neutron condensate does not interact with the proton vortex in the absence of entrainment,
so that there are no terms in equation~(\ref{force}), depending on ${\pmb V}_{\rm sn}-{\pmb V}_{\rm L}$
(a subsequent calculation confirms this expectation).
In equation~(\ref{force}) 
$D$, $D^\prime$, and $D_z$ are the kinetic coefficients to be determined below 
(the coefficients $D$ and $D'$
have already been introduced in Section \ref{intro}).
In what follows we assume that the difference 
${\pmb u}_{\rm e}-{\pmb V}_{\rm L}$ 
is sufficiently small and restrict ourselves to calculations valid 
in linear order in
${\pmb u}_{\rm e}-{\pmb V}_{\rm L}$.
In this approximation the coefficients $D$, $D^\prime$, and $D_z$ are velocity-independent
and to determine them we, for simplicity, 
consider two cases. 
First, assume that the vector ${\pmb u}_{\rm e}-{\pmb V}_{\rm L}$ 
is collinear with ${\pmb e}_z$.
Then only the last term $\propto D_z$ survives in equation (\ref{force})
and
the corresponding force is directed along the vortex line.
But any such force should vanish 
since the magnetic field cannot scatter electrons traveling along the axis $z$.
We come to conclusion that $D_z=0$.
Assume now that the vector ${\pmb u}_{\rm e}-{\pmb V}_{\rm L}$ 
lies in the $xy$-plane
(see Fig.~\ref{vortvel}).
Then equation~(\ref{force}) can be conveniently represented as
\begin{align}
&{\pmb F}_{{\rm npe}\rightarrow {\rm V}}=
D ({\pmb u}_{\rm e}-{\pmb V}_{\rm L})
+D'[{\pmb e}_z\times ({\pmb u}_{\rm e}-{\pmb V}_{\rm L})].
&
\label{force2}
\end{align}
%

Our system is assumed to be stationary in the coordinate system moving with the vortex, 
which means that the force 
${\pmb F}_{{\rm npe}\rightarrow {\rm V}}$ must be balanced
by some `external' force, ${\pmb F}_{\rm ext}$, acting on a vortex
(in NSs it can be, for example, the tension and/or buoyancy forces, which do not depend on particle velocities; 
see equation (\ref{forcebalance0}) in Section \ref{intro} 
and \citealt{dg17} for details),
\begin{align}
&{\pmb F}_{{\rm npe}\rightarrow {\rm V}}+{\pmb F}_{\rm ext}=0.&
\label{forcebalance}
\end{align}

Let us express ${\pmb F}_{{\rm npe}\rightarrow {\rm V}}$ through 
the stress tensor of npe-matter far from the vortex.
With this aim we make use of the total momentum conservation 
(see Appendix \ref{subtle} for more details)
\begin{align}
&\frac{\partial {\pmb G}}{\partial t} + 
\nabla_i \Pi_{ik} = {\pmb f}_{\rm ext},&
\label{momentum}
\end{align}
where ${\pmb G}$ and $\Pi_{ik}$ are the momentum density 
and stress tensor for npe-matter (including the electromagnetic field), respectively, and
${\pmb f}_{\rm ext}$ is the external force density localized in the vortex core.%
%
\footnote{In fact, this requirement is not necessary for derivation of equation (\ref{cond1}) below.}
%
Now, let us choose a cylinder of unit length 
with the symmetry axis coinciding with the vortex line,
and radius $r_0$ satisfying the inequality
$l\gg d_{\rm B} \gg r_0 \gg \lambda$ (see Fig.\ \ref{vortvel}).
Integrating equation~(\ref{momentum}) over the cylinder volume 
and using the Gauss theorem, 
as well as the fact that the system is stationary, $\partial {\pmb G}/\partial t=0$,
and that, by definition, $\int {\pmb f}_{\rm ext} \, {\rm d}V={\pmb F}_{\rm ext}$,
one finds
\begin{align}
&
-\oint \Pi_{ik}\, n_k {\rm d} S +{\pmb F}_{\rm ext}=0,
&
\label{momentum2}
\end{align}
or, taking into account the relation (\ref{forcebalance}),
\begin{align}
&
{\pmb F}_{{\rm npe}\rightarrow {\rm V}}=-\oint \Pi_{ik}\, n_k \, {\rm d} S,
&
\label{cond1}
\end{align}
where the integration is performed over the cylinder surface;
${\rm d} S$ is the surface element; and $n_k$ is the outer normal unit vector.

Formula (\ref{cond1}) is very useful since it allow us to find
the force on the vortex provided that the stress tensor far from the vortex
is known (an explicit expression for $\Pi_{ik}$ is presented in Appendix \ref{subtle}). 
However, far from the vortex (at $r \gg \lambda$) the vortex magnetic field 
and the proton superfluid velocity, generated by the vortex, 
are exponentially suppressed (see equation \ref{B} and, e.g., \citealt{degennes99}).
As it is shown in Appendix~\ref{subtle},
in these circumstances only electrons contribute to
the integral (\ref{cond1}).%
%
\footnote{This statement is not precise; see the text below 
and Appendix \ref{subtle} for a detailed explanation.}
%
The electron stress tensor $\Pi^{({\rm e})}_{ik}$ is given by the standard expression
(see, e.g., \citealt{ll81}),
\begin{align}
&
\Pi^{({\rm e})}_{ik}= \sum_{{\pmb p} \sigma} p_i 
v_k
n_{\pmb p\sigma},
&
\label{Piik}
\end{align}
where ${\pmb p}$, $\epsilon_{\pmb p}$, 
and ${\pmb v}=\partial \epsilon_{\pmb p}/\partial {\pmb p}$ 
are the electron (kinetic) momentum, energy, and velocity, respectively; 
$n_{\pmb p\sigma}$ is the electron distribution function; and 
summation is assumed over the electron momenta and spins.
If $n_{\pmb p\sigma}$ does not explicitly depend on spins (our case), one has
$\sum_{\pmb p\sigma} \equiv 2/(2 \pi \hbar)^3 \int d^3 {\pmb p}$.

The problem, therefore, reduces to finding the electron distribution function 
far from the vortex.
At distances $l\gg r \gg \lambda$ it can be written 
as a sum of three terms
to be discussed below: 
\begin{align}
&
n_{\pmb p \sigma}=n_{\pmb p \sigma }^{({\rm eq})}+\delta n_{\pmb p \sigma}^{({\rm sc})}+\delta n_{\pmb p \sigma}^{({\rm ind})}.
&
\label{np1}
\end{align}
The first term here represents the incident flow of electrons with velocity ${\pmb u}_{\rm e}$. 
In the coordinate system in which ${\pmb V}_{\rm L}=0$
it is given by the shifted Fermi-Dirac distribution function,
\begin{align}
&
n_{\pmb p \sigma }^{({\rm eq})} = n_{{\pmb p}0}(\epsilon_{\pmb p}-\mu_{\rm e}-{\pmb p}[{\pmb u}_{\rm e}-{\pmb V}_{\rm L}]),
&
\label{npeq}
\end{align}
where $n_{{\pmb p}0}(\epsilon_{\pmb p})\equiv 1/[{\rm e}^{\epsilon_{\pmb p}/T}+1]$ 
and $\mu_e$ is the electron chemical potential far from the vortex.
Clearly, this term does not contribute to the force (\ref{cond1}), 
since 
it
`does not know' about the presence of the vortex line.

The second term in equation~(\ref{np1}) describes the electrons scattered by the vortex.
The asymptotic expression for $\delta n_{\pmb p \sigma}^{({\rm sc})}$, 
valid at large distances from the vortex,
has been
derived by \cite{sonin76,gs76,aggk81}, and is given by (see also Appendix \ref{distrib})
\begin{align}
&
\delta n_{\pmb p \sigma}^{({\rm sc})}(r,\,\theta, \, p,\, \theta_p) &=& 
\frac{{\rm d} n_{{\pmb p}0}}{{\rm d} \epsilon_{\pmb p}} 
\left\{ ({\pmb u}_{\rm e}-{\pmb V}_{\rm L}){\pmb p} \, \sigma_{\rm \|}
\right.
&
\nonumber\\
&
&+& \left. [{\pmb e}_z \times ({\pmb u}_{\rm e}-{\pmb V}_{\rm L})]{\pmb p} \, \sigma_{\rm \perp} \right\}
\, \frac{\delta(\theta_{p} - \theta)}{r},
&
\label{NSC}
\end{align}
where $\sigma_{\rm \|}$ is the well-known transport cross-section,
\begin{align}
&
\sigma_{\|} = \int_{-\pi}^\pi  \sigma(\gamma)\, (1- {\rm cos \, \gamma}) \,{\rm d}\gamma
&
\label{sigma_par}
\end{align}
and 
\begin{align}
&
\sigma_{\perp} = \int_{-\pi}^\pi  \sigma(\gamma) \, {\rm sin} \, \gamma \,{\rm d}\gamma.
&
\label{sigma_perp}
\end{align}

In equations~(\ref{NSC})--(\ref{sigma_perp}) $r$ and $\theta$ 
are, respectively, the cylindrical radius and angle -- 
the coordinates of a point
in the cylindrical coordinate system with the centre at the vortex line (see Fig.\ \ref{scheme});
$\theta_p$ is the angle coordinate of momentum ${\pmb p}$ in the same 
coordinate system; 
$\sigma(\gamma)$ is the effective differential cross-section 
for scattering of electrons off the vortex line; 
$\gamma$ is the scattering angle: 
$\gamma=\theta_{p_{\ci f}}-\theta_{p_{\ci i}}$,
where ${\pmb p}_{\ci i}$ and ${\pmb p}_{\ci f}$ are
the electron momenta before and after scattering, respectively (see also Fig.\ \ref{geometry}).%
%
\footnote{
Note that, apparently, \cite{sonin76,gs76,aggk81} define $\gamma$ (and hence $\sigma_\perp$) with an opposite sign.
As a result, terms depending on $\sigma_\perp$ in our equations (\ref{NSC}) and (\ref{D'})
differ by the sign from the corresponding equations in these references.}
%
Because of
2d-character of our scattering problem,
the cross-sections (\ref{sigma_par}) and (\ref{sigma_perp})
have a dimension of length.

It is easy to understand the general structure of the expression (\ref{NSC}).
The correction $\delta n_{\pmb p \sigma}^{({\rm sc})}$, 
describing scattered electrons,
is proportional to ${\rm d} n_{{\pmb p}0}/{\rm d}\epsilon_{\pmb p}$,
which means that only electrons close to the Fermi surface can scatter off the vortex line 
-- an expected result for a strongly degenerate matter;
far from the vortex  $\delta n_{\pmb p \sigma}^{({\rm sc})}\propto 1/r$,
which is natural since the total number of scattered electrons should be conserved;
delta-function in equation~(\ref{NSC}) indicates that scattered electrons move radially from the vortex 
to the observation point ($\theta_p=\theta$);
finally, the combinations $({\pmb u}_{\rm e}-{\pmb V}_{\rm L}){\pmb p}$
and $[{\pmb e}_z \times ({\pmb u}_{\rm e}-{\pmb V}_{\rm L})]{\pmb p}$ in curly brackets in equation~(\ref{NSC})
are the only scalars that can be composed of the available vectors in the problem.

Now let us turn to the third term in equation~(\ref{np1}).
It describes a subtle effect, ignored so far in the literature, 
and related to the fact that 
scattered electrons carry a charge.
As a result, a weak electric field will appear far from the vortex
and this will slightly change the electron distribution function. 
In addition, this will 
also modify the proton chemical potential.
In Appendix \ref{subtle} we show that contribution to 
${\pmb F}_{{\rm npe}\rightarrow {\rm V}}$ from both these effects 
mutually cancel each other,
so we should not care about the last term in equation~(\ref{np1}).  

Correspondingly, the final expression for the force (\ref{cond1}) can be rewritten as
\begin{align}
&
{\pmb F}_{{\rm npe}\rightarrow {\rm V}} =-\oint \delta\Pi^{({\rm e, \, sc})}_{ik}\, n_k \, {\rm d} S,
&
\label{COND2}
\end{align}
where 
\begin{align}
&
\delta \Pi^{({\rm e, \, sc})}_{ik} =\sum_{{\pmb p} \sigma} p_i \,  
v_k
\, \delta n^{({\rm sc})}_{\pmb p\sigma}.
&
\label{dPi}
\end{align}
Integrating (\ref{COND2}) using equations~(\ref{NSC}) and (\ref{dPi}), 
one arrives at the expression (\ref{force2}) for ${\pmb F}_{{\rm npe}\rightarrow {\rm V}}$,
in which
\begin{align}
&
D= \frac{1}{2} \sum_{{\pmb p}\sigma} \left(-\frac{{\rm d} n_{{\pmb p}0}}{{\rm d} \epsilon_{\pmb p}}\right)
\, p_{\perp}^2 \, v_{\perp} \sigma_{\|}(p_{\perp}),
&
\label{D}\\
&D'= \frac{1}{2} \sum_{{\pmb p}\sigma} \left(-\frac{{\rm d} n_{{\pmb p}0}}{{\rm d} \epsilon_{\pmb p}}\right)
\, p_{\perp}^2 \, v_{\perp} \sigma_{\perp}(p_{\perp}).
&
\label{D'}
\end{align}
The expressions (\ref{D}) and (\ref{D'}) were derived long ago by \cite{sonin76} (see also \citealt{gs76,aggk81,sonin16}). 
In equations~(\ref{D}) and (\ref{D'}) $p_\perp$ and $v_\perp$ are, respectively, 
the projections of the electron momentum and 
velocity on the plane $xy$. 
To determine the coefficients $D$ and $D^\prime$ we need 
to calculate the cross-sections $\sigma_{\|}$ and $\sigma_{\perp}$,
which are, generally, the functions of $p_\perp$.
The next section is devoted to such calculation.

\section{Cross-sections $\sigma_{\|}$ and $\sigma_{\perp}$ due to electron scattering 
	off the vortex magnetic field}
\label{cross_section}

\subsection{Simple calculation within the classical scattering theory}
\label{classic}

Let us first calculate the cross-sections 
$\sigma_{\|}$ and $\sigma_{\perp}$ assuming that 
the electrons can be treated classically.
This is a justified assumption provided that their 
wavelength, $\hbar/p$, is much smaller than the typical lengthscale 
of the magnetic field variation, $\lambda$.
Indeed, it is well known \citep{ll80} that in the latter case one can 
use the standard Boltzmann equation (\ref{boltzmann}) 
with the external Lorentz force incorporated, 
and this result is independent of whether
the electrons in the system are degenerate or not. 
The Boltzmann equation can (in principle) be solved by the characteristics method,
and the asymptotic correction $\delta n_{{\pmb p}\sigma}^{({\rm sc})}$ 
for the distribution function $n_{{\pmb p\sigma}}$
can be derived, 
which is equivalent to finding the cross-sections $\sigma_\|$ and $\sigma_\perp$ (see equation \ref{NSC}).
On the other hand, the same Boltzmann equation can be used to find $\sigma_\|$ and $\sigma_\perp$
for purely classic problem of particle scattering on the magnetic field of a vortex. 
It is clear, therefore, that any `classic' derivation
should give the correct answer 
for the cross-sections
$\sigma_\|$ and $\sigma_\perp$.
This conclusion is additionally verified in Section \ref{quantum} by a more rigorous 
calculation within the quasiclassical scattering theory.
%
\begin{figure}
	\begin{center}
		\includegraphics[width=0.3\textwidth]{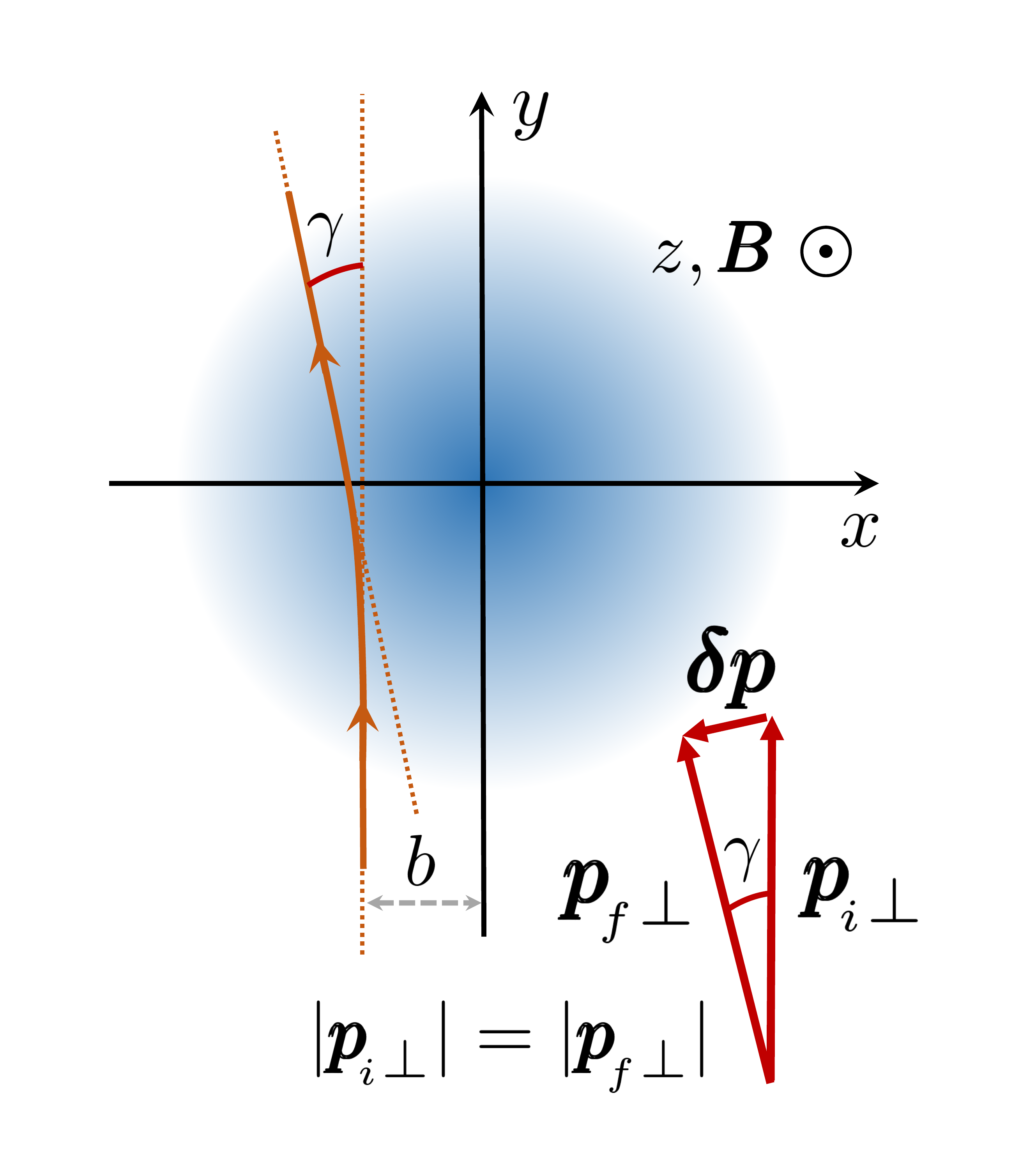}
		\caption{\label{geometry}
Electron trajectory (schematic) in the magnetic field of a vortex.
A projection ${\pmb p}_{i\perp}$ of an incident electron 
is directed along the axis $y$; 
${\pmb p}_{f\perp}$ is the momentum projection of the scattered electron,
$\gamma$ is the scattering angle,
$b$ is the impact parameter.
		}
	\end{center}
\end{figure}
%

Assume that the projection ${\pmb p}_{\ci i\perp}$ of the momentum ${\pmb p}_{\ci i}$ 
of an incident electron on the $xy$-plane is directed along the axis $y$,
while the electron impact parameter is $b$ (see Fig.~\ref{geometry}).
Electrons are fast, so that the transferred momentum 
$\delta {\pmb p}={\pmb p}_{\ci f}-{\pmb p}_{\ci i}={\pmb p}_{\ci f\perp}-{\pmb p}_{\ci i\perp}$ 
(due to the action of the Lorentz force) 
and the scattering angle $\gamma$
are both small. 
In these conditions $\delta {\pmb p}$ is parallel to the axis $x$ and one may approximately write  
\begin{align}
&\pmb{\delta  p} = \int \frac{e_{\rm e}}{c} \, {\pmb v}_\perp \times {\pmb B}(r) \, {\rm d}t 
\approx \pmb{e}_x \int \frac{e_{\rm e}}{c} \,  B(\sqrt{b^2+y^2}) \, {\rm d}y,&
\label{dp}
\end{align}
where 
in the second equality we use the fact that 
${\pmb v}_{\perp} \approx ({\rm d}y/{\rm d}t) \, {\pmb e}_y$ 
(${\pmb e}_x$ and ${\pmb e}_y$ are the unit vectors along the axes $x$ and $y$).
The scattering angle $\gamma(b)$ is now given by
\begin{align}
&
\gamma(b) \approx \frac{\delta p}{p_{\perp}} =\frac{1}{p_{\perp}} \,
\int_{-\infty}^{+\infty} \frac{e_{\rm e}}{c} \, B(\sqrt{b^2+y^2}) \, {\rm d}y.
&
\label{gamma}
\end{align}
The differential cross-section $\sigma(\gamma)$ is defined by the formula (e.g., \citealt{sonin16}): 
$\sigma(\gamma)={\rm d}b(\gamma)/{\rm d}\gamma$.
Plugging this definition into equations~(\ref{sigma_par}) and (\ref{sigma_perp}) 
and using the fact that $\gamma$ is small, 
one arrives at the following expressions 
for $\sigma_{\perp}$~and~$\sigma_{\rm \|}$,
\begin{align}
&
\sigma_{\|} 
\approx \int_{-\infty}^{+\infty} \frac{\gamma(b)^2}{2} \, {\rm d}b,
&
\label{sigma_par2}\\
&
\sigma_{\perp} 
\approx \int_{-\infty}^{+\infty} \, \gamma(b) \, {\rm d} b,&
&
\label{sigma_perp2}
\end{align}
where $\gamma(b)$ is given by equation~(\ref{gamma}).
Formula (\ref{sigma_perp2}) for $\sigma_\perp$ can be easily integrated 
for an {\it arbitrary} vortex magnetic field ${\pmb B}$, 
\begin{align}
&\sigma_\perp = \frac{e_{\rm e} \Phi}{p_\perp c} \approx 3 \,{\rm fm}\, \left(\frac{e_{\rm e}}{e_{\rm p}}\right)
\left(\frac{\Phi}{\Phi_0}\right)\left(\frac{1 \, {\rm fm}^{-1}}{k_\perp}\right),&
\label{sigma_perp3}
\end{align}
where $k_\perp \equiv p_\perp/\hbar$.
To obtain equation (\ref{sigma_perp3}) we make use of the fact 
that the total magnetic flux carried by the vortex is
$\Phi = \int B \, {\rm d}b {\rm d}y$. 
The cross-section $\sigma_\perp$ is negative for electrons because 
the magnetic field turns them to the left (negative angles $\gamma$).

\begin{figure}
	\begin{center}
		\includegraphics[width=0.3\textwidth]{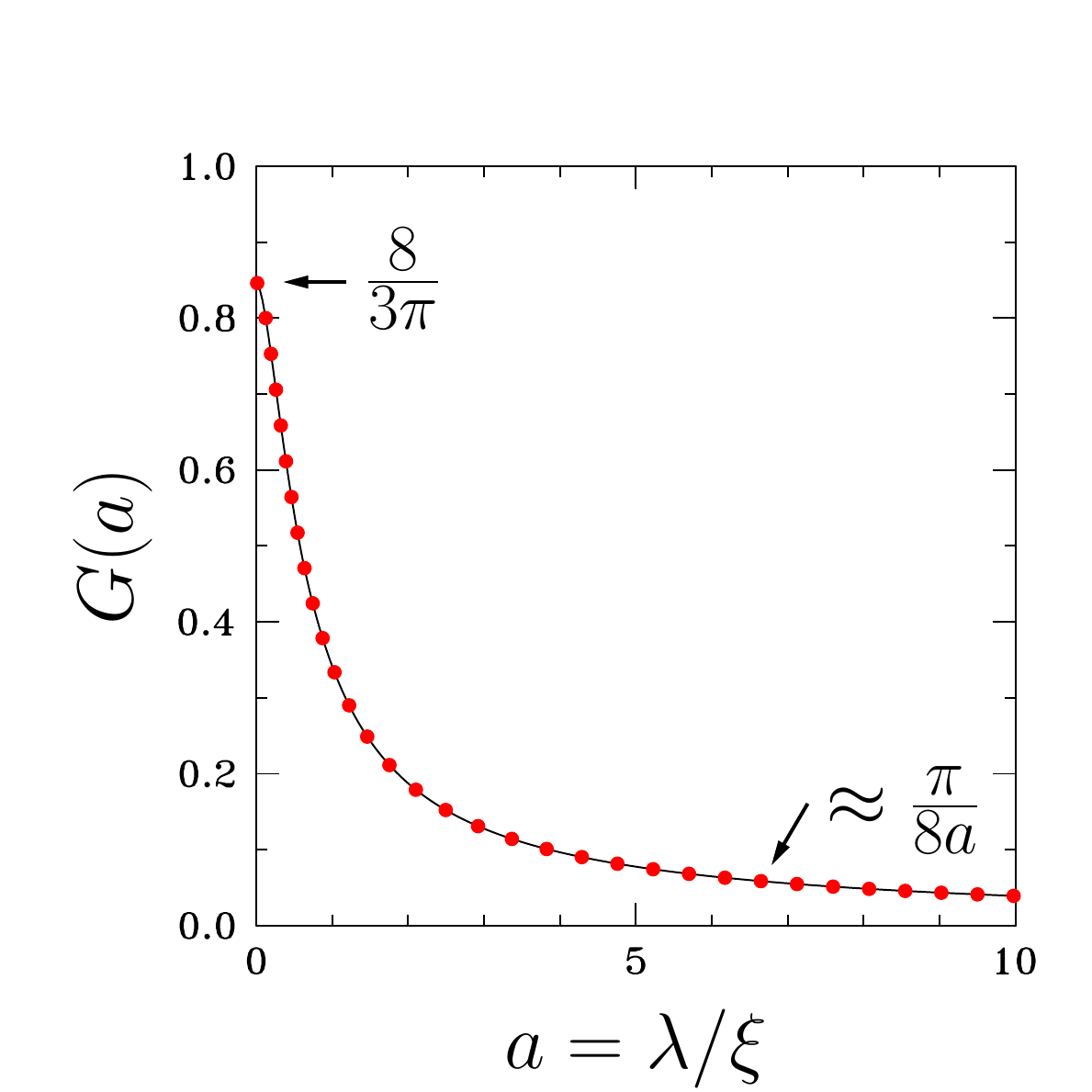}
		\caption{\label{Fa}
			The function $G(a)$. Solid line: fitting formula (\ref{Fapprox});
			circles: numerical calculation using equation (\ref{F}).
		}
	\end{center}
\end{figure}
In turn, equation~(\ref{sigma_par2}) for $\sigma_{\|}$ can be represented
in a more `quantum-mechanical' form (see Appendix \ref{integral} for details),
\begin{align}
&\sigma_{\|} 
= \frac{e_{\rm e}^2}{4 \pi p_{\perp}^2 c^2}\,
\int_{-\infty}^{+\infty} |\Pi(q)|^2 \, {\rm d}q,&
\label{SIGMA_PAR3}
\end{align}
where the form-factor $\Pi(q)$ equals
\begin{align}
&\Pi(q)=\int B(\sqrt{b^2+y^2}) \, {\rm e}^{-i q b} \, {\rm d}b {\rm d}y.&
\label{form}
\end{align}
For the model (\ref{B}) of the vortex magnetic field, we have
$\Pi(q)=2 \Phi J_1(q \xi)/[q \xi (1+q^2 \lambda^2)]$, so that 
$\sigma_{\|}$ can finally be presented as
\begin{align}
&\sigma_{\|}=\frac{e_{\rm e}^2 \Phi^2 }{\pi p_{\perp}^2 c^2 \xi}\, \, G(\lambda/\xi)
=\left(\frac{e_{\rm e}}{e_{\rm p}}\right) \left(\frac{\Phi}{\Phi_0}\right)
\left(\frac{1}{k_{\rm Fe}\xi}\right) G(\lambda/\xi) \, \sigma_\perp,&
\label{sigma_par4}
\end{align}
where the function $G(a)$ is plotted in Fig.~\ref{Fa} and is given by
\begin{align}
&
G(a) = \int_{-\infty}^{\infty} \frac{J_1(x)^2}{x^2 (1+x^2 a^2)^2} \, {\rm d}x.
&
\label{F}
\end{align}
In two limiting cases, $a\gg 1$ and $a\ll 1$, this function equals, respectively, $\pi/(8 a)$
and $8/(3 \pi)$.
We fit $G(a)$ by the following approximate formula, which reproduces the results
of our numerical calculations with the maximum error $\sim 0.3\%$,
\begin{align}
&
G(a)=\frac{8+c_1 a + c_2 a^2 + c_3 a^3+ (3 \pi^2/8) c_4 a^4}{3 \pi (1+c_5 a^3+c_4 a^5)},
&
\label{Fapprox}
\end{align}
where 
$c_1 = -0.9411$; 
$c_2 = -28.679$;
$c_3 = -1.7340$;
$c_4 = 12.774$;
$c_5 = -6.3427$.

\subsection{Quasiclassical calculation using the scattering theory}
\label{quantum}

Here our aim is to determine the cross-sections $\sigma_{\|}$ and $\sigma_\perp$
using the standard scattering theory. 
Because the electron wavelength, $\hbar/p$, is much smaller than other typical lengthscales
in the problem, the scattering can be considered  quasiclassically.
For nonrelativistic electrons the problem of quasiclassical electron scattering by a vortex 
was analyzed long ago by \cite{cleary68}, 
who, however, employed a different model of the vortex magnetic field 
(see also \citealt{sonin16} for a recent discussion of this problem);
in Section~\ref{nonrel} we present a similar analysis.
An extension of the results of Section~\ref{nonrel} 
to the relativistic case is given in Section~\ref{rel}.

\subsubsection{Nonrelativistic electrons}
\label{nonrel}

The electron wave function $\Psi({\pmb r})$ in the presence 
of a proton vortex with the magnetic field ${\pmb B}$
can be found from the stationary Schr$\ddot{\rm o}$edinger equation,
\begin{align}
&
E \Psi =\frac{1}{2 m_{\rm e}} \left(
-\imath \hbar \,{\pmb \nabla} - \frac{e_{\rm e}}{c} {\pmb A}
\right)^2 \Psi
-\hat{\pmb \mu} {\pmb B} \, \Psi, 
&
\label{schrodinger}
\end{align}
where $E$ and $m_{\rm e}$ are the electron energy and bare mass, respectively; 
$\hat{\pmb \mu}$ is the electron magnetic momentum operator (see, e.g., \citealt{ll77});
and ${\pmb A}$ 
is the vector-potential of 
the electromagnetic field given by the equation (\ref{A}).%
%
\footnote{The electrostatic potential $\phi$ produced by the slightly inhomogeneous matter in the very
vicinity of the vortex line is small and can be neglected in the equation (\ref{schrodinger}).}
%
In what follows we shall ignore the last term in equation (\ref{schrodinger}),
describing interaction of the electron spin with the magnetic field.
It can be shown (by essentially repeating the derivation that will be presented below) 
that the contribution from this term to $\sigma_{\|}$ and $\sigma_\perp$ is negligible
for scattering of unpolarized quasiclassical electron.
To simplify formulas we also assume that the (conserved) 
$z$-component of the electron momentum vanishes.

We are interested in the solution to equation (\ref{schrodinger}) 
with the asymptotic behavior 
\begin{align}
&
\Psi = {\rm e}^{i k_\perp y }+\frac{f(\theta)}{\sqrt{r}} \, {\rm e}^{i k_\perp r},
&
\label{Psiasym}
\end{align}
describing the incident plane wave moving along the axis $y$ 
(the first term)
and the scattered cylindrical wave (the second term) 
(see, e.g., \citealt{cleary68,ll77,sonin16}).
In equation (\ref{Psiasym}) $k_\perp=p_\perp/\hbar$; 
$r$ and $\theta$ are the coordinates introduced in Section \ref{vort};
and $f(\theta)$ is the scattering amplitude,
which is related to the differential cross-section $\sigma(\gamma)$ by the formula 
(e.g., \citealt{ll77}; note that in this section the scattering angle $\gamma$ coincides 
with the angular coordinate~$\theta$):
\begin{align}
&
\sigma = |f(\theta)|^2.
&
\label{diff}
\end{align}

Generally, $\Psi(r,\, \theta)$ can be decomposed as
\begin{align}
&
\Psi(r,\, \theta) = \sum_{l=-\infty}^{+\infty} \, {\rm e}^{il\theta} \, Q_l(r).
&
\label{Psidecomp}
\end{align}
Plugging this expression into equation (\ref{schrodinger}) (with the last term omitted) 
and making use of equation (\ref{A}),
one finds
\begin{align}
&
\frac{1}{r}\frac{\rm d}{{\rm d}r}\left(r \frac{{\rm d}Q_l}{{\rm d}r}\right)
+\left[
k_\perp^2-\left(\frac{l}{r}-\frac{e_{\rm e}}{\hbar c}\, A_{\theta}\right)^2
\right] \, Q_l=0,
&
\label{Ql}
\end{align}
In the system without a vortex ($A_\theta=0$) 
a solution to this equation, regular at $r=0$, 
is the Bessel function, $J_{|l|}(k_\perp r)$.
At $k_\perp r \gg |l^2-1/4|$ its asymptote is
\begin{align}
&
J_{|l|}(k_\perp r) \approx \sqrt{\frac{2}{\pi k_\perp r}} \, 
{\rm cos} \, \left(k_\perp r - \frac{\pi}{2}|l|-\frac{\pi}{4}\right).
&
\label{Jlasym}
\end{align}
On the other hand, at $A_\theta \neq 0$ the function
$Q_l$ will have an asymptote which differ from (\ref{Jlasym}) only 
by a phase shift $\delta_l$ and normalization factor $C_l$, i.e.,
\begin{align}
&
Q_l = C_l \, \sqrt{\frac{2}{\pi k_\perp r}} \, 
{\rm cos} \, \left(k_\perp r - \frac{\pi}{2}|l|-\frac{\pi}{4} + \delta_l \right).
&
\label{Qlasym}
\end{align}
Using equations (\ref{Psidecomp}) and (\ref{Qlasym}) and requiring
that the asymptotic expression for $\Psi$ takes the asymptotic form (\ref{Psiasym}),
one arrives at the following expression for the normalization factor 
$C_l$ and the scattering amplitude $f(\theta)$ \citep{cleary68,sonin16}:
\begin{align}
&
C_l={\rm e}^{i\frac{\pi|l|}{2}+i \delta_l},
&
\label{Cl}\\
&
f(\theta) =\sqrt{\frac{1}{2 \pi i k_\perp}} \, \sum_{l=-\infty}^{\infty} \, {\rm e}^{i l \theta} 
\left({\rm e}^{2 i \delta_l}-1\right).
&
\label{f}
\end{align}
Now, plugging equations (\ref{diff}) and (\ref{f}) 
into the definitions (\ref{sigma_par}) and (\ref{sigma_perp})
and recalling that $\gamma=\theta$,
one obtains the following expressions for the cross-sections $\sigma_{\|}$ 
and $\sigma_\perp$, first derived by \cite{cleary68} and confirmed subsequently by \cite{op85,nh95,sonin97,shelankov98,shelankov00,sonin16}:
\begin{align}
&
\sigma_{\|}=\frac{2}{k_\perp} \, \sum_{l=-\infty}^{\infty} \, {\rm sin}^2(\delta_{l+1}-\delta_l),
&
\label{sigma_par6}\\
&
\sigma_{\perp}=\frac{1}{k_\perp} \, \sum_{l=-\infty}^{\infty} \, {\rm sin}(2\delta_{l}-2\delta_{l+1}).
&
\label{sigma_perp6}
\end{align}

Both these quantities depend on the phase shifts $\delta_l$.
To calculate $\delta_l$, we make use of the fact that 
for electrons with $k_\perp \lambda \gg 1$ one may work in the quasiclassical (WKB) approximation.%
%
\footnote{To solve equation (\ref{Ql}) in the quasiclassical approximation one needs
first to transform it to the form of the standard one-dimensional Shr\"odinger equation by
introducing a new function, $\chi_l(r) \equiv \sqrt{r} \, Q_l(r)$ \citep{ll77}. }
%
In the quasiclassical approximation the main contribution to the scattering amplitude $f(\theta)$
and to the differential cross-section $\sigma$
comes from the partial waves with large $l$ \citep{ll77}, i.e., 
large angular momenta.
For such $l$ one can present $\delta_l$ as \citep{cleary68,ll77}
\begin{align}
&
\delta_l \approx \int_{|l|/k_\perp}^\infty \, \frac{l \, e_{\rm e} A_{\theta}}{\hbar c \, r \, 
\sqrt{k_{\perp}^2-l^2/r^2}} \, {\rm d} r 
&
\nonumber\\
&
\,\,\,\;=- \frac{e_{\rm e}}{e_{\rm p}} \, \frac{l \zeta}{k_{\perp}}\, 
\int_{|l|/k_\perp}^\infty \, \frac{\mathcal{P}(r)}{ r \, 
\sqrt{r^2-l^2/k_\perp^2}} \, {\rm d} r,
&
\label{dl}
\end{align}
where the function $\mathcal{P}(r)$ is defined by equation (\ref{A}).
Using the expression (\ref{dl}) one can easily calculate the cross-sections $\sigma_\|$ and $\sigma_\perp$
from the equations (\ref{sigma_par6}) and (\ref{sigma_perp6}).
The calculation can be simplified by noticing that $\delta_l$ is a slowly varying function of $l$,
so that one may treat $l$ as a continuous variable and write $\delta_{l+1}-\delta_l\approx {\rm d}\delta_l/{\rm d}l$.
Then equations (\ref{sigma_par6}) and (\ref{sigma_perp6}) can be presented as \citep{cleary68}
\begin{align}
&
\sigma_{\|}\approx \frac{2}{k_\perp} \, \int_{l=-\infty}^{\infty} \,\left(\frac{{\rm d}\delta_l}{{\rm d}l}\right)^2 \, {\rm d}l,
&
\label{sigma_par7}\\
&
\sigma_{\perp}\approx -\frac{2}{k_\perp} \, \int_{l=-\infty}^{\infty} \,\frac{{\rm d}\delta_l}{{\rm d}l} \, {\rm d}l.
&
\label{sigma_perp7}
\end{align}
Although these formulas look different from those obtained in Section \ref{classic},
they are, in fact, completely equivalent to the expressions (\ref{sigma_par2}) and (\ref{sigma_perp2})
for $\sigma_\|$  and $\sigma_\perp$, as shown in Appendix \ref{equiv}.
Correspondingly, in the limit $k_\perp \lambda \gg 1$ one should use 
the cross-sections $\sigma_\|$ and $\sigma_\perp$,
given by the formulas (\ref{sigma_perp3}) and (\ref{sigma_par4}).

It is interesting to note that $\sigma_\|$ and $\sigma_\perp$, calculated 
in the limit $k_\perp \lambda \gg 1$, 
differ drastically from 
those obtained in the opposite limit, $k_\perp \lambda \ll 1$, 
corresponding to the classic Aharonov-Bohm effect 
(when the flux tube can be treated as infinitely thin; \citealt{ab59}). 
It is easy to show that in the limit $k_\perp \lambda \ll 1$
the phase shifts are given by the formula (e.g., \citealt{sonin16})
\begin{align}
&
\delta_l=\left(|l|-\left|l+ \frac{e_{\rm e}}{e_{\rm p}}\zeta \right| \right)\, \frac{\pi}{2},
&
\label{dl3}
\end{align}
and hence from equations (\ref{sigma_par6}) and (\ref{sigma_perp6}) it follows that 
\begin{align}
&
\sigma_\|=\frac{2}{k_\perp} \, {\rm sin}^2 \left(\frac{\pi e_{\rm e}}{e_{\rm p}}\zeta \right)
=\frac{2}{k_\perp} \,{\rm sin}^2 \left(\frac{e_{\rm e}}{e_{\rm p}}\,\frac{\pi \Phi}{2\Phi_0} \right),
&
\label{sigma_par8}\\
&
\sigma_\perp=\frac{1}{k_\perp} \, {\rm sin}\left( \frac{2 \pi e_{\rm e}}{e_{\rm p}}\zeta \right)
=\frac{1}{k_\perp} \, {\rm sin} \left(\frac{e_{\rm e}}{e_{\rm p}}\,\frac{\pi \Phi}{\Phi_0} \right).
&
\label{sigma_perp8}
\end{align}
The cross-section $\sigma_{\|}$ in this limit was (implicitly) calculated, e.g., in \cite{op85,as10};
in turn, $\sigma_\perp$ was calculated, e.g., in \cite{sonin97,sonin16}
and implicitly considered in \cite{op85,nh95,shelankov98,shelankov00}. 
Note that for a proton vortex $\sigma_\|=2/k_\perp$ and $\sigma_\perp=0$
because in that case $\Phi=\Phi_0$.

\subsubsection{Relativistic generalization}
\label{rel}

The results of the previous section cannot be used directly since electrons 
in the internal layers of NSs are ultrarelativistic (e.g., \citealt*{hpy07}).
In order to find
the cross-sections $\sigma_\|$ and $\sigma_\perp$ 
for a relativistic electron 
one needs, in principle, to solve the scattering problem for the Dirac equation.
A similar problem was studied long ago by \cite{aw89}, 
who calculated the differential cross-section $\sigma$
for the scattering of a relativistic fermion off a 
vortex (cosmic string). 
Although these authors were interested in finding $\sigma$ 
in the limit of infinitely thin vortex, $k_\perp \lambda \ll 1$, 
their approach can also be used in our situation.
Namely, \cite{aw89} exploited the symmetry of the vortex under $z$ translations.
It turns out that for $z$-independent problems there exists a representation 
of the Dirac $\gamma$-matrices that allow one to decouple the Dirac equation
into two independent first-order differential equations%
%
\footnote{One equation is for `spin-up' and one for `spin-down' electron.} 
%
for two-component spinors \citep{devega78}.
Each of these equations is {\it exactly} equivalent 
to the nonrelativistic Shr\"odinger equation (\ref{Ql}).
Therefore, all the consideration of Section \ref{nonrel} remains unaffected 
and leads to the same 
cross-sections, $\sigma_\|$ and $\sigma_\perp$, 
as in the nonrelativistic limit.
We refer the interested reader to the work by \cite{aw89} for more details.

\section{Results and comparison with the previous works}
\label{disc}

Using equations (\ref{sigma_perp3}) and (\ref{sigma_par4})
we can now calculate the coefficients $D$ and $D'$ 
from the formulas (\ref{D}) and (\ref{D'}):
\begin{align}
&
D = \frac{e_{\rm e}^2 \Phi^2 p_{\rm Fe}^2}{8 \pi^2 \hbar^3 c^2 \xi} \, G(\lambda/\xi)
&
\nonumber\\
&
\,\,\,\;=\frac{3 \pi}{8} \left(\frac{e_{\rm e}}{e_{\rm p}}\right) \left(\frac{\Phi}{\Phi_0}\right)
\left(\frac{1}{k_{\rm Fe}\xi}\right) G(\lambda/\xi)  D',
&
\label{D2}\\
&
D' = \frac{e_{\rm e}}{c} \, n_{\rm e} \Phi
=\pi \hbar \, n_{\rm e} \left(\frac{e_{\rm e}}{e_{\rm p}}\right)
\left(\frac{\Phi}{\Phi_0}\right),
&
\label{D'2}
\end{align}
where 
$n_{\rm e}=p_{\rm Fe}^3/(3 \pi^2 \hbar^3)$.
One sees that $D>0$, 
which means that the (dissipative) longitudinal force on a vortex, 
${\pmb F}_\| \equiv D ({\pmb u}_{\rm e}-{\pmb V}_{\rm L})$, 
acts in the direction of the axis $y$ (see equation \ref{force2}).
This is an expected result since the momentum of electrons along the axis $y$ 
decreases in the course of scattering (see Fig.\ \ref{geometry}).  
In turn, $D'<0$, i.e., the transverse force on a vortex,
${\pmb F}_\perp \equiv D' [{\pmb e}_z\times({\pmb u}_{\rm e}-{\pmb V}_{\rm L})]$,
acts in the direction of the axis $x$.
This result is also reasonable, since 
exactly the same force (Lorentz force) acts on electrons in the opposite direction. 

One may note that ${\pmb F}_\perp$ formally coincides with the so called Magnus force%
%
\footnote{This coincidence takes place only in the quasiclassical limit, $k_{\rm Fe}\lambda \gg 1$.}
%
${\pmb F}_{\rm M}$ (see equation \ref{Fmagnus} and recall the quasineutrality condition, $n_{\rm e}=n_{\rm p}$),
which is the transverse force acting on a vortex from superconducting protons
and is well-defined for extreme type-II superconductors (when $\xi \ll \lambda$).
To understand this coincidence, assume for a moment that $\xi \ll \lambda$ for our problem
and recall that the force ${\pmb F}_{\rm npe \rightarrow V}={\pmb F}_\|+{\pmb F}_\perp$,
calculated by us above, is the
{\it total} force from the npe-mixture on a vortex.
What are the actual mechanism and an actual particle species 
participating in transferring 
the momentum
to the vortex core we have not yet discussed.
Clearly,
this cannot be neutrons, because they 
in no way interact with the vortex.
Also, this cannot be electrons because they, generally,
scatter off the magnetic field localized {\it far} from the vortex core ($\lambda \gg \xi$);
see Appendix \ref{Fm} for a more detailed justification of this statement.
This magnetic field is generated and supported by the superconducting proton currents,
consequently, scattered electrons transfer their momentum to superconducting proton component,
but not to the vortex.
We come to conclusion that in this example only protons are able to transfer 
the momentum directly to the vortex core.
How does it happen?
The mechanism of the transverse force appearance is essentially the same
as in liquid helium-II (e.g., \citealt{sonin87,sonin16}).
This is because the proton superfluid velocity,
generated by the vortex, 
scales as $\propto 1/r$
at distances $r$ from the vortex centre, 
such that $\xi \ll r \ll \lambda$ \citep{nv66,ll80,degennes99}.
The superfluid velocity near the vortex core in helium-II behaves in exactly the same way
and this is known to produce a transverse force on a vortex 
if  
a superfluid transport current 
is applied to the system (e.g., \citealt{sonin87,sonin16}).
This force can be found from equations (\ref{sfl}) and (\ref{mom})
in Appendix \ref{subtle}
by considering a momentum
carried 
by
superconducting protons 
per unit time
through the walls of a cylinder of radius $r$,
with the result 
that it equals ${\pmb F}_{\rm M}$ \citep{nv66,sonin87,sonin16}.  
Therefore, it is not surprising that ${\pmb F}_\perp={\pmb F}_{\rm M}$
if $\xi\ll \lambda$.
But the expression for the force ${\pmb F}_\perp$
does not 
contain 
$\xi$ and/or $\lambda$, so this result should remain unchanged
in the more general case of arbitrary ratio between these parameters. 

Note that our results for the transverse force ${\pmb F}_\perp$
agree with the assumptions about the form of the force made, e.g., by \cite{as10,gas11}
and disagree with the conclusions of \cite{jones91,jones06}, 
where it is argued that ${\pmb F}_\perp=0$.
The coefficient $D'$ would vanish only in the limit of large electron wavelength 
(when $\sigma_\perp=0$, see equation \ref{sigma_perp8}),
but this limit is not realized in NSs, for which $\hbar/p_{\rm Fe}\ll \lambda$.
For completeness, below we provide the coefficients $D$ and $D'$ calculated in the limit
$k_{\rm Fe} \lambda \ll 1$
(using $\sigma_\|$ and $\sigma_\perp$ given, respectively, 
by equations \ref{sigma_par8} and \ref{sigma_perp8}):
\begin{align}
&
D=2 \hbar \, n_{\rm e} \, {\rm sin}^2 \left(\frac{e_{\rm e}}{e_{\rm p}} \frac{\pi \Phi}{2\Phi_0}\right),
&
\label{D4}\\
&
D'=\hbar \, n_{\rm e} \, {\rm sin} \left(\frac{e_{\rm e}}{e_{\rm p}} \frac{\pi \Phi}{\Phi_0}\right).
&
\label{D'4}
\end{align}
Similar expression for the coefficient $D$ was obtained in \cite{op85,nh95,as10},
while the coefficient $D'$ in this limit was studied in 
\cite{op85,nh95,sonin97,shelankov98,shelankov00,sonin16}.

Now, let us discuss in some more detail the longitudinal force ${\pmb F}_\|$
on a vortex and compare it with the results available in the literature.
The longitudinal force due to electron scattering off the vortex magnetic field 
was calculated by 
\cite{jones87} (see also \citealt*{hrs86}) within classical mechanics.
Using the formulas of Section \ref{classic}, 
it is easy to verify that our result (equation \ref{D2}) agrees with that of Jones.
The force ${\pmb F}_\|$ was also calculated by  
\cite{als84}
(see also \citealt*{sss82}). 
Strictly speaking, \cite{als84} considered a bit different problem, namely, 
the electron scattering off the neutron vortices, which can carry magnetic field 
due to the entrainment effect \citep{ab76}.
However, their solution can easily be applied to our problem (\citealt{ss95}).
\cite{als84} used a very different method of derivation of ${\pmb F}_\|$
and, moreover, worked in the Born approximation.%
%
\footnote{\cite{als84} did not find the transverse force ${\pmb F}_\perp$,
which vanishes in the Born approximation, 
because in that case $\sigma(-\gamma)=\sigma(\gamma)$ and thus $\sigma_\perp=0$.}
%
Meanwhile, this approximation is unjustified 
for relatively large magnetic fluxes, associated with the vortex, $\Phi \sim \Phi_0$, 
for which it can lead to incorrect results.%
%
\footnote{
It is worth noting that the Born approximation can be inadequate 
even for $\Phi \ll \Phi_0$. 
This is the case 
for a nonrelativistic problem of electron scattering off an infinitely thin flux tube,
if the latter is treated with the Schr\"odinger equation (see, e.g., \citealt{aall84}).
However, the same problem, analyzed in the Born approximation 
making use of the Dirac equation, 
gives correct asymptotic expression for the differential cross-section, 
valid at $\Phi \ll \Phi_0$ (see \citealt{vs90} for details).
 }
%
Thus, it is interesting to look whether our force ${\pmb F}_\|$ differs from that of \cite{als84}. 
 
Actually, \cite{als84} calculated the so called `velocity coupling time 
between the plasma and the core superfluid', $\tau_{\rm v}$.
In the limit of $k_{\rm Fe}\xi \gg 1$ it is given by%
%
\footnote{This formula follows from equation (30b) of \cite{als84}.
Note a misprint in equation (30b):
instead of $x^2+\alpha^2$ there should be $x^2/2+\alpha^2$,
as we checked by independent calculation of $\sigma_\|$ in the Born approximation.}
%
%
\begin{align}
&
\frac{1}{\tau_{\rm v}}=\frac{3}{2} \frac{p_{\rm Fe} c}{m_{\rm p}c^2}\, 
\frac{1}{\alpha \tau_0} \, G(\lambda/\xi),
&
\label{tv}
\end{align}
where $\alpha=2 p_{\rm Fe}\xi/\hbar$; $\tau_0^{-1}= \pi N_\tau \Phi^2$;
the function $G(\lambda/\xi)$ is defined by equation (\ref{F});
and
\begin{align}
&
N_\tau = \frac{2 \pi}{\hbar} \, n_{\rm v} \, \left( \frac{e_{\rm e} \hbar}{2 m_{\rm e} c}\right)^2\,
\left(\frac{m_{\rm e}c^2}{p_{\rm Fe}c}\right)^2\, \frac{p_{\rm Fe} c}{(\pi \hbar c)^2}
&
\label{Ntau}
\end{align}
with $n_{\rm v}$ being the surface vortex density.
As shown, e.g., in \cite{ss95,asc06,kg17}, 
this relaxation time is related to the force 
on a vortex per unit length by the formula:
\begin{align}
&
{\pmb F}_{\rm Alpar\, \|}=\frac{m_{\rm p}n_{\rm p}}{n_{\rm v} \tau_{\rm v}} 
\, ({\pmb u}_{\rm e}-{\pmb V}_{\rm L}).
&
\label{Falpar}
\end{align}
%
\begin{figure}
	\begin{center}
		\includegraphics[width=0.5\textwidth]{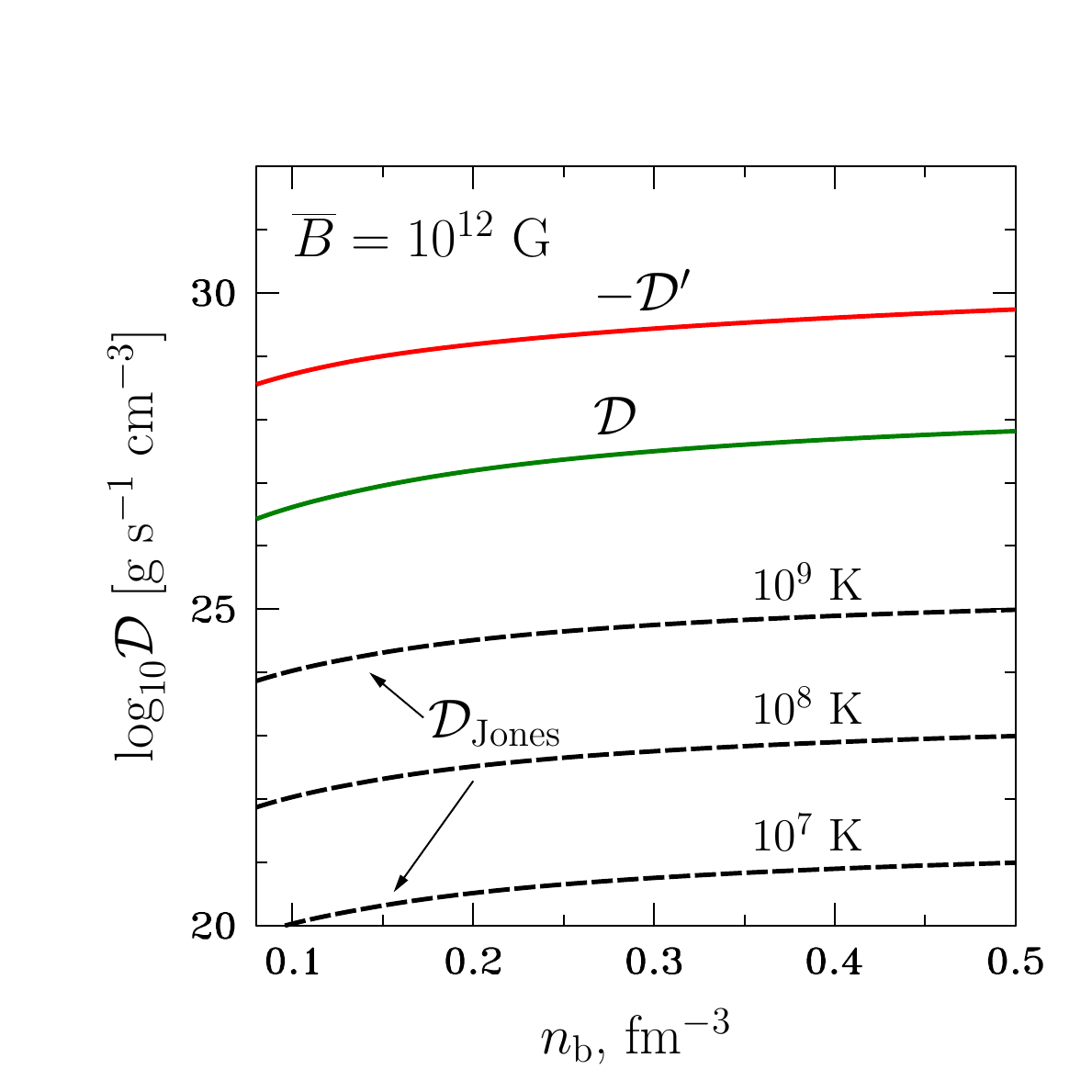}
		\caption{\label{DD'}
			Coefficients $\mathcal{D}$, $-\mathcal{D}'$, and $\mathcal{D}_{\rm Jones}$
			versus $n_{\rm b}$ for $\overline{B}=10^{12}$~G. 
			The temperature-dependent coefficient $\mathcal{D}_{\rm Jones}$
			is plotted for three stellar temperature: $T=10^7$, $10^8$, and $10^9$~K.
		}
	\end{center}
\end{figure}
%
Plugging equation (\ref{tv}) into (\ref{Falpar})
and comparing the coefficient at $({\pmb u}_{\rm e}-{\pmb V}_{\rm L})$
with the expression (\ref{D2}),
one verifies that, somewhat unexpectedly, 
${\pmb F}_{\rm Alpar \, \|}= {\pmb F}_\|$, i.e.,
the longitudinal force calculated by \cite{als84}
exactly coincides with our result.
 
Equations (\ref{D2}) and (\ref{D'2}) determine the force on a single vortex.
However, in astrophysical applications one is usually interested 
in the force density $\pmb{ \mathcal{F}}_{\rm npe\rightarrow V}$ acting on a system of vortices. 
Assuming that we have a locally rectilinear array of proton vortices 
with the surface density $n_{\rm v} = \overline{B}/\Phi$, 
one can present $\pmb {\mathcal{F}}_{\rm npe\rightarrow V}$ as (cf. equation \ref{force})
\begin{align}
&
\pmb{\mathcal{F}}_{\rm npe\rightarrow V}=-\mathcal{D} [{\pmb e}_z \times [{\pmb e}_z\times({\pmb u}_e-{\pmb V}_{\rm L})]]
+\mathcal{D}'[{\pmb e}_z\times ({\pmb u}_{\rm e}-{\pmb V}_{\rm L})],
&
\label{F2}
\end{align}
where 
\begin{align}
&
\mathcal{D}=n_{\rm v} \, D = 
\frac{3\pi}{8} \frac{e_{\rm e}}{e_{\rm p}} \left(\frac{\Phi}{\Phi_0}\right) 
\left(\frac{1}{k_{\rm Fe}\xi}\right) \,G(\lambda/\xi) \,\, \mathcal{D}',
&
\label{D3}\\
&
\mathcal{D}'=n_{\rm v} \, D' 
=\frac{e_{\rm e}}{c} \, n_{\rm e}\, \overline{B}.
&
\label{D3'}
\end{align}
One sees that 
\begin{align}
&
\frac{\mathcal{D}}{\mathcal{D}'}=\frac{D}{D'} \approx
\begin{dcases}
\frac{3 \pi^2}{64}\left(\frac{1}{k_{\rm Fe}\lambda}\right)\approx
0.46 \left(\frac{1}{k_{\rm Fe}\lambda}\right), & \quad \lambda\gtrsim \xi,\\
\frac{1}{k_{\rm Fe}\xi}, & \quad \lambda\ll \xi.
\end{dcases}
&
\label{DD'ratio}
\end{align}
(The latter case is relevant for neutron vortices and is not interesting for us here.)
Generally, $\mathcal{D}$ and $D$ are
much smaller than, respectively, $\mathcal{D}'$ and $D'$ for typical NS conditions, 
for which $k_{\rm Fe} \lambda \sim (30-50)$ (see Fig.\ \ref{param}).%
%
\footnote{
\label{15}
Note that the ratio $D/D'$ was denoted as $\mathcal{R}$ in \cite{gagl15} 
and was estimated to be $\mathcal{R}\sim 1.9 \times 10^{-4}$ 
(see their equation 71 for a more accurate expression).
This estimate disagrees with 
our result (\ref{DD'ratio}): $D/D' \sim 0.46/50\sim 9 \times 10^{-3}$,
leading to $D/(D'\mathcal{R}) \sim 50$.
Correspondingly, the magnetic field evolution timescales in \cite{gagl15,dg17}
should be revised (Gusakov et al., in preparation).
}
%
This is also illustrated in Fig.\ \ref{DD'}, where the coefficients
$\mathcal{D}$ and $\mathcal{D}'$ are plotted as functions of $n_{\rm b}$
for $\overline{B}=10^{12}$~G. 
For comparison, we also present
the dissipative coefficient
$\mathcal{D}_{\rm Jones}$ used by  \cite{jones91,jones06} and, recently, by \cite{blb17}
in their studies of the magnetic field expulsion timescale from the NS cores.
(Note, that these authors completely ignored the effect 
of electron scattering by the vortex magnetic field, thus assuming
$\mathcal{D}= \mathcal{D}_{\rm Jones}$, $\mathcal{D}'=0$.)
The temperature-dependent coefficient $\mathcal{D}_{\rm Jones}$ 
enters the expression for
the dissipative force, similar to the first term in equation (\ref{F2}).
This force arises due to 
the electron scattering off the unpaired proton quasiparticles
localized in the vortex core.
\cite{jones91,jones06} and \cite{blb17} used 
the following simple order-of-magnitude estimate for this coefficient 
(see also \citealt{gs76}):
\begin{align}
&
\mathcal{D}_{\rm Jones}\approx \frac{n_{\rm e} p_{\rm Fe}}{c \tau_{\rm ep}} 
\, \left(\frac{\overline{B}}{2 H_{\rm c2}} \right),
&
\label{Djones}
\end{align}
where $H_{\rm c2}=\Phi/(2 \pi \xi^2)$ and $\tau_{\rm ep}$
is the typical timescale of electron-proton collisions 
in the normal (nonsuperfluid and nonsuperconducting) matter. 
It is given by the formula (see, e.g., \citealt{ys91a,ys91b}): 
$\tau_{\rm ep} = p_{\rm Fe} n_{\rm e}/(J_{\rm ep} c)$,
where the friction coefficient $J_{\rm ep}$ is 
\begin{align}
&
J_{\rm ep} = 2 \times 10^{28}  \left(\frac{T}{10^8\,{\rm K}}\right)^2 \left(\frac{\rho_0}{\rho}\right)^{5/3} 
\left(\frac{n_{\rm e}}{n_0}\right)^{4/3} \frac{\rm g}{\rm cm^3 \, s}
&
\label{Jep}
\end{align}
and $\rho_0=2.8 \times 10^{14}$~g cm$^{-3}$.
The coefficient $\mathcal{D}_{\rm Jones}$ in Fig.\ \ref{DD'}
is plotted for three stellar temperatures, $T=10^7$, $10^8$, and $10^9$~K.
One sees that $\mathcal{D}_{\rm Jones}$ is always small in comparison to $\mathcal{D}$.
This result is independent of the magnetic induction $\overline{B}$, since 
both these coefficients are proportional to $\overline{B}$.
Thus, \cite{jones06} and \cite{blb17} substantially underestimate
the typical timescales of magnetic field evolution in NSs.

\subsection*{Inclusion of muons}
\label{muons}

If there are muons in the system, they will also scatter off the proton vortices.
The corresponding analogue of equation (\ref{force}) for the force on a vortex 
in the case of npe$\mu$-matter is 
(we already set to zero the component of the force along the axis $z$):
\begin{align}
&{\pmb F}_{{\rm npe\mu}\rightarrow {\rm V}}=
-D [{\pmb e}_z \times [{\pmb e}_z\times({\pmb u}_e-{\pmb V}_{\rm L})]]
+D'[{\pmb e}_z\times ({\pmb u}_{\rm e}-{\pmb V}_{\rm L})]
&
\nonumber\\
&
-D_\mu [{\pmb e}_z \times [{\pmb e}_z\times({\pmb u}_\mu-{\pmb V}_{\rm L})]]
+D'_\mu[{\pmb e}_z\times ({\pmb u}_{\rm \mu}-{\pmb V}_{\rm L})],
&
\label{force3}
\end{align}
where ${\pmb u}_{\mu}$ is the muon velocity far from the vortex; $D_\mu$ and $D'_\mu$
are the muon coefficients similar to, respectively, $D$ and $D'$.
In order to write the force in the form (\ref{force3}) we have already used the screening 
condition (\citealt{jones91,jones06,gas11,gd16}), 
which, in the presence of muons (and neglecting entrainment), can be written as
\begin{align}
&e_{\rm p} n_{\rm p} {\pmb V}_{\rm sp} 
+e_{\rm e} n_{\rm e} {\pmb u}_{\rm e}
+
e_{\mu} n_{\mu} {\pmb u}_{\mu}=0.
&
\label{screening2}
\end{align}
Here and below
$e_{\mu}$, $n_{\mu}$, $p_{\rm F\mu}$, and $k_{\rm F\mu}$ are the muon charge, 
number density, Fermi momentum and Fermi wave number, respectively.
Assuming now $k_{\rm F\mu}\lambda \gg 1$ 
(which is a valid assumption sufficiently far from the threshold for the muon appearance), 
it is straightforward to show that
the coefficients $D_\mu$ and $D'_\mu$  
are given by the same equations 
(\ref{D2}) and (\ref{D'2}) as for electrons,
with the obvious replacements
 $e_{\rm e} \leftrightharpoons e_\mu$,
$p_{\rm Fe} \leftrightharpoons p_{\rm F\mu}$,
and $n_{\rm e} \leftrightharpoons n_\mu$.
Equations (\ref{D3}) and (\ref{D3'}) can be adjusted to allow for muons in a similar way.

\section{Conclusions}
\label{concl}

We calculated the force acting on 
a proton vortex from neutron-proton-electron 
mixture at vanishing stellar temperature and neglecting, for simplicity, 
entrainment effects between the superfluid neutrons and superconducting protons.
It was assumed that 
far from the vortex the electron and proton charge current densities 
are non-zero and equal to one another because of the screening condition (\ref{screening}).
The force is found by analyzing the outgoing momentum flow,
generated by the vortex per unit time. 
This approach has been previously used in application to liquid helium-II 
and electrons in type-II superconductors, e.g., by \cite{sonin76,gs76,aggk81}.
Our main results are summarized as follows:

\begin{itemize}

\item 
For typical NS conditions the electron wavelength is much smaller
than all other relevant lengthscales in the problem, 
in particular, than the London penetration depth, $\lambda$.
This permits us to use a quasiclassical scattering theory in order to determine 
the electron cross-sections $\sigma_\|$ and $\sigma_\perp$, responsible 
for the appearance of transverse ${\pmb F}_\perp$ and longitudinal ${\pmb F}_\|$ 
forces on a vortex.
In Section \ref{classic} it is shown that purely classic calculation of 
$\sigma_\|$ and $\sigma_\perp$ gives the same result. 
Note that the celebrated differential Aharonov-Bohm cross-section \citep{ab59}
cannot be used to calculate the force on a vortex in our situation
(as it is done, e.g., in application to quark matter by \citealt{as10}),
because it is obtained in the opposite limit, $k_{\rm Fe}\lambda\ll 1$.%
%
\footnote{A dissipative force ${\pmb F}_\|$ 
on a color-magnetic flux tube in quark matter can be easily calculated 
in the limit of small wavelength  
of scattered particles, $k_{\rm F}\lambda \gg 1$, following our approach.
The resulting expression will be suppressed by a factor of $k_{\rm F}\lambda$ 
in comparison to the result presented in \cite{as10}. }
%

\item
The calculated transverse force ${\pmb F}_\perp$ coincides (only in the limit $k_{\rm Fe}\lambda \gg 1$)
with the ordinary Magnus force, 
discussed in the context of superconductors, e.g., by \cite{nv66,kopnin02}.
It also equals to the (minus) Lorentz force acting on electrons in the magnetic field of a vortex.
This result proves that the assumptions 
made, e.g., in \cite{as10,gas11} about the form of ${\pmb F}_\perp$ are correct. 
At the same time, our result disagrees with the conclusion of \cite{jones91,jones06} that ${\pmb F}_\perp=0$.

\item
The longitudinal force on a proton vortex ${\pmb F}_\|$ 
is, typically, smaller than ${\pmb F}_\perp$ 
by a factor of $k_{\rm Fe} \lambda\sim (30 - 50)$.
It coincides 
with the force calculated by \cite{jones87} 
and (after some straightforward adjustment) 
with the force on a neutron vortex calculated by \cite{als84}.
The latter coincidence is rather surprising since \cite{als84} worked within the Born approximation,
which is not justified in our problem.

\item
\cite{jones06} and \cite{blb17} ignored in their analysis electron scattering
by the magnetic field of a vortex, thus effectively setting 
$D'=0$.
Instead, 
they considered a different scattering mechanism, 
namely, scattering of electrons off the proton localized excitations in the vortex core.
As is shown in Section \ref{disc}, this mechanism leads to 
a longitudinal force much smaller than our ${\pmb F}_\|$ 
(i.e., the friction coefficient $D_{\rm Jones}$, suggested by \citealt{jones06},
is much smaller than our coefficient $D$ given by equation \ref{D2} and can be ignored).
This means that \cite{jones06} and \cite{blb17} substantially underestimate
the typical timescale $\tau_B$ for the magnetic field evolution in the NS core.

\item 
In Section \ref{disc} we show how our results should be modified to allow for muons in the system.

\end{itemize}

The results obtained in this paper confirm the form of the force on a vortex 
postulated, e.g., by \cite{as10} and \cite{gas11}.
As a consequence, simple estimates of the (very long) magnetic field evolution timescales
made in \cite{gagl15,dg17} and numerically found in \cite{eprgv15} 
look more realistic 
than those obtained in \cite{jones06,blb17} (but see footnote \ref{15}).
These estimates imply, however, that NS matter as a whole is immobile, 
the assumption that can be incorrect for magnetized NSs (\citealt*{gko17}; \citealt{og18}).
Account for macroscopic fluid motions in the core may dramatically accelerate the
magnetic field evolution. 

This work can be extended in a number of ways.
First, it is straightforward to account for additional particle species in the system (e.g., hyperons)
and allow for non-vanishing entrainment between the superfluid baryons.
Secondly, the 
approach developed in the present paper can be directly applied to 
study forces that act on a neutron vortex. 
However, we do not expect
that the expression for such force
will differ noticeably from that already used in the literature (e.g., \citealt{gas11}). 
Thirdly, it would be very interesting to 
generalize the results obtained here to the case of finite stellar temperatures.
This can be a more difficult task since at finite $T$ one should also 
account for scattering of neutron and proton thermal Bogoliubov excitations 
off the quasiparticles localized in the vortex core \citep{kopnin02,sonin16}.
Whether additional forces appearing due to such scattering
play a role in the NS dynamics 
remains an open question to be investigated in the future.

\section*{Acknowledgments}

I am grateful to E.M.~Kantor, D.G.~Yakovlev, and A.I.~Chugunov for numerous useful discussions.
I would also like to thank 
E.B.~Sonin, E.M.~Kantor, and V.A.~Dommes for a critical reading
of the draft version of this paper and valuable comments.
This work is supported in part by the Foundation 
for the Advancement of Theoretical Physics and Mathematics `BASIS'
(grant No.~17-12-204-1) and by RFBR (grant No.~19-52-12013).

\appendix 

\section{Derivation of the asymptotic distribution function (\ref{NSC}) for scattered electrons  }
\label{distrib}

\begin{figure}
	\begin{center}
		\includegraphics[width=0.3\textwidth]{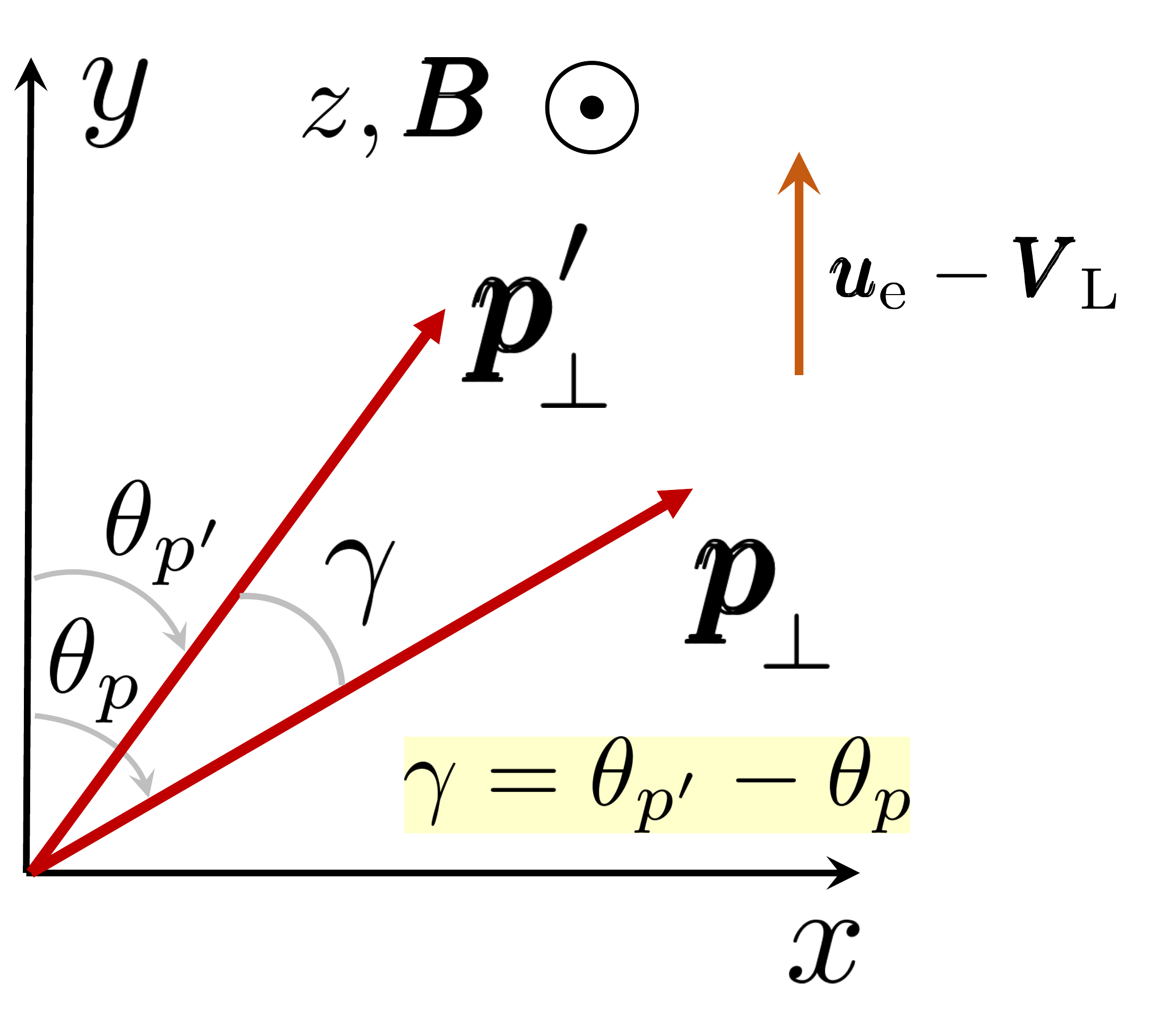}
		\caption{\label{cherteg}
			Scattering geometry for electrons. ${\pmb p}_\perp$ and ${\pmb p}'_\perp$
			are the projections of electron momenta ${\pmb p}$ and ${\pmb p}'$ on the $xy$-plane.
			See the text for details.
		}
	\end{center}
\end{figure}

Let us first calculate 
the change ${\rm d} \mathcal{N}_{\pmb p\sigma}$ in the
number of electrons
with momentum ${\pmb p}$ and spin $\sigma$ per unit time
due to electron scattering
by the vortex in the linear approximation 
in ${\pmb u}_{\rm e}-{\pmb V}_{\rm L}$.%
%
\footnote{Note that it is not a real scattering in a statistical sense, 
since there are no `element of chance' in the problem (\citealt{gl87}): the wave 
function of an electron in the field of a vortex is a well defined quantity 
that can be found from the Schr\"odinger (or Dirac) equation. }
%
Since scattering by the magnetic field of a vortex is elastic, 
the $z$-component $p_z$ of the momentum ${\pmb p}$ and the absolute
value $p_\perp$ of the projection of ${\pmb p}$ on the plane $xy$
are both conserved. 
Also, as noted in Section \ref{nonrel} 
the effect of magnetic field interaction with the electron spins is small and can be ignored;
thus, the spin is also a conserved quantity. 
The effect of scattering, therefore, reduces to changing the electron angle coordinate from $\theta_p$
to $\theta_{p'}$ (see Fig.\ \ref{cherteg} showing the geometry of the problem).
In these circumstances ${\rm d}\mathcal{N}_{\pmb p\sigma}$ can be written as
\begin{align}
&
{\rm d}\mathcal{N}_{\pmb p\sigma} = -{\rm d}^3{\pmb p}\, {\rm d} z \, \int_{-\pi}^{\pi} n_{\pmb p \sigma}^{({\rm eq})}\,
 v_{{\pmb p}\perp} \sigma(\gamma, p_\perp) 
\, 
{\rm d}\theta_{p'} 
&
\nonumber\\
&
\quad\quad\,\,
+{\rm d}^3{\pmb p} \, {\rm d} z\, \int_{-\pi}^{\pi} n_{\pmb p' \sigma}^{({\rm eq})} \,
v_{{\pmb p}'\perp} 
\sigma(-\gamma, p_{\perp}') \, {\rm d}\theta_{p'} ,
&
\label{dN}
\end{align}
where the first term in the right-hand side represents electrons 
scattered from $\theta_{p}$ to some $\theta_{p'}$ 
(correspondingly, the scattering angle is $\gamma=\theta_{p'}-\theta_{p}$),
while the second term represents 
inverse process, $\theta_{p'}\rightarrow \theta_{p}$
(the corresponding scattering angle equals~$-\gamma$).
In equation (\ref{dN}) $v_{{\pmb p}\perp}=v_{{\pmb p'}\perp}$ 
is the electron velocity in the $xy$-plane;
$\sigma(\gamma, \, p_\perp)$ is the standard 2D differential cross-section, which has 
a dimension of length (see, e.g., \citealt{ll77} and section~\ref{cross_section} for details).
Further, $n_{\pmb p\sigma}^{({\rm eq})}$ is the equilibrium distribution function (\ref{npeq}),
unperturbed by the vortex. 
It is justifiable to use $n_{\pmb p \sigma}^{({\rm eq})}$ 
in equation (\ref{dN}) since we work in the linear approximation in the velocities.
%
%
Substituting
\begin{align}
& n_{\pmb p \sigma}^{({\rm eq})} \approx
n_{\pmb p 0}(\epsilon_{\pmb p})-\frac{{\rm d}n_{\pmb p0}}{{\rm d} \epsilon_{\pmb p}} {\pmb p}({\pmb u}_e-{\pmb V}_{\rm L})
&
\nonumber\\
&
\quad\quad
=n_{\pmb p 0}(\epsilon_{\pmb p})-\frac{{\rm d}n_{\pmb p0}}{{\rm d} \epsilon_{\pmb p}} p |{\pmb u}_e-{\pmb V}_{\rm L}| 
\, {\rm cos }\theta_p,&
\nonumber\\
& n_{\pmb p' \sigma}^{({\rm eq})} \approx
n_{\pmb p 0}(\epsilon_{\pmb p})-\frac{{\rm d}n_{\pmb p0}}{{\rm d} \epsilon_{\pmb p}} {\pmb p}'({\pmb u}_e-{\pmb V}_{\rm L})
\nonumber\\
&
\quad\quad
=n_{\pmb p 0}(\epsilon_{\pmb p})-\frac{{\rm d}n_{\pmb p0}}{{\rm d} \epsilon_{\pmb p}} p |{\pmb u}_e-{\pmb V}_{\rm L}| 
\, {\rm cos} \theta_{p'}&
\label{dnps}
\end{align}
into (\ref{dN}), and using the relation $\theta_{p'}=\gamma+\theta_{p}$, 
we arrive at the formula
\begin{align}
&
{\rm d} \mathcal{ N}_{{\pmb p}\sigma} = {\rm d}^3{\pmb p}\, {\rm d} z 
\, \frac{{\rm d} n_{{\pmb p}0}}{{\rm d} \epsilon_{\pmb p}} 
\left\{ ({\pmb u}_e-{\pmb V}_{\rm L}){\pmb p} \, \,\sigma_{\rm \|}
\right.&
\nonumber\\
&
\left.
\quad\quad\,\,
+ [{\pmb e}_z \times ({\pmb u}_e-{\pmb V}_{\rm L})]{\pmb p} \, \sigma_{\rm \perp} \right\},
&
\label{dNp}
\end{align}
where the naturally appearing cross-sections $\sigma_{\|}$ and $\sigma_{\perp}$
are given by equations (\ref{sigma_par}) and (\ref{sigma_perp}), respectively.
The quantity 
${\rm d} \mathcal{ N}_{{\pmb p}\sigma}/{\rm d} z$
describes 
the total number of electrons with momentum ${\pmb p}$ and spin $\sigma$ 
produced
in the vicinity of a vortex line
per unit time and per unit vortex length
due to
scattering  
by the magnetic field of a vortex.
It is this quantity which is responsible for a deviation 
$\delta n_{\pmb p \sigma}^{({\rm sc})}(r,\,\theta, \, p,\, \theta_p)$ 
(see equation \ref{NSC})
of the electron distribution function from
$n_{{\pmb p}\sigma}^{({\rm eq})}$ far from the vortex.
To find $\delta n_{\pmb p \sigma}^{({\rm sc})}$ one needs to solve 
the kinetic Boltzmann equation, which has the asymptotic form
\begin{align}
&
{\pmb v}_{\pmb p} \cdot {\pmb \nabla} \delta n_{{\pmb p} \sigma}^{({\rm sc})} = {\rm St} \, n_{\pmb p\sigma},
&
\label{kinetic}
\end{align}
where ${\rm St} \, n_{\pmb p\sigma}$ is the `collision integral'
given by 
\begin{align}
&
{\rm St} \, n_{\pmb p\sigma} = 
\frac{\delta(r)}{2 \pi r}\,\,
\frac{{\rm d} \mathcal{ N}_{{\pmb p}\sigma}}{ {\rm d}^3{\pmb p}\, {\rm d} z }.
&
\label{Stnp}
\end{align}
The delta-function here indicates that the scattering occurs in the very vicinity of
the vortex (where exactly is not important since we are interested in the asymptotic solution 
for $\delta n_{\pmb p \sigma}^{({\rm sc})}$ at $r\gg \lambda$);
the normalization factor $2 \pi r$ in the denominator ensures that the total number 
of scattered electrons is  
${\rm d}^3{\pmb p} \, \int{\rm  St} \, n_{{\pmb p}\sigma} \, {\rm d}V  
= {\rm d} \mathcal{ N}_{{\pmb p}\sigma}$.
The equation (\ref{kinetic}) can be rewritten as
${\pmb \nabla} \cdot ({\pmb v}_{\pmb p} \, \delta n_{{\pmb p} \sigma}^{({\rm sc})} )
={\rm St} \, n_{\pmb p\sigma}$; the solution to this equation can be readily obtained and coincides 
with the expression~(\ref{NSC}).

\section{Stress tensor for npe-matter and proof of equation~(\ref{COND2})   }
\label{subtle}

\subsection{Stress tensor}

At $T=0$ the (nonrelativistic) equations of motion for protons far from the vortex 
consist of 
the continuity equation,
\begin{align}
&\frac{\partial n_{\rm p}}{\partial t}+{\rm div} \,(n_{\rm p} {\pmb V}_{{\rm sp}})=0,&
\label{cont}
\end{align}
superfluid equation (see, e.g., \citealt{nv66,putterman74,aggk81})
\begin{align}
&\frac{\partial {\pmb V}_{{\rm sp}}}{\partial t}=\frac{e_{\rm p}}{m_{\rm p}} \, {\pmb E} - {\pmb \nabla}
\left(\frac{\mu_{\rm p}}{m_p}+\frac{1}{2} V_{{\rm sp}}^2 \right),&
\label{sfl}
\end{align}
and the condition (\citealt{ll80,degennes99})
\begin{align}
&{\pmb \nabla}\times{\pmb V}_{\rm sp}=-\frac{e_{\rm p}}{m_{\rm p} c} \, {\pmb B},&
\label{B-u}
\end{align}
specific to superconducting systems.
In equations~(\ref{cont})--(\ref{B-u}) $n_{\rm p}$ and $m_{\rm p}$ are the proton number density and mass, 
respectively;
$\mu_{\rm p}$ is the proton chemical potential, 
defined in the coordinate system, 
in which ${\pmb V}_{{\rm sp}}=0$. 
From these equations one can derive the proton momentum conservation equation,
\begin{align}
&\frac{ \partial{\pmb G}_{\rm p}}{\partial t} = - \partial_k \left( m_{\rm p} n_{\rm p} \, V_{{\rm sp}\, i} V_{{\rm sp} \, k} \right)
- 
n_{\rm p} \, {\pmb \nabla} \mu_{\rm p} 
&
\nonumber\\
&
\quad\quad\,
+ e_{\rm p} n_{\rm p}  
\, {\pmb E} +\frac{e_{\rm p}}{c} \, n_{\rm p} \, {\pmb V}_{{\rm sp}}\times {\pmb B},&
\label{mom}
\end{align}
where ${\pmb G}_{\rm p}=m_{\rm p} n_{\rm p} {\pmb V}_{\rm sp}$ is the proton momentum density.
A similar equation can also be written out for neutrons,
\begin{align}
&\frac{ \partial{\pmb G}_{\rm n}}{\partial t} = 
- \partial_k \left( m_{\rm n} n_{\rm n} \, V_{{\rm sn}\, i} V_{{\rm sn} \, k} \right)
- 
n_{\rm n} \, {\pmb \nabla} \mu_{\rm n}.&
\label{mom_n}
\end{align}
where ${\pmb G}_{\rm n}=m_{\rm n} n_{\rm n} {\pmb V}_{\rm sn}$; 
$n_{\rm n}$, $m_n$, and $\mu_n$ are the neutron number density, mass, and chemical potential,
respectively.

Now we turn to 
the momentum conservation equation for electrons.
Since $k_{\rm Fe}\lambda \gg 1$
it can be derived from the standard Boltzmann kinetic equation,
\begin{align}
&
\frac{\partial n_{\pmb p\sigma}}{\partial t}+{\pmb v}\, 
\frac{\partial n_{\pmb p\sigma}}{\partial {\pmb r}}
+ e_{\rm e} \left(
{\pmb E} + \frac{1}{c} \, {\pmb v}\times {\pmb B}
\right)\frac{\partial n_{\pmb p \sigma}}{\partial{\pmb p}}=0.
&
\label{boltzmann}
\end{align}
Multiplying it by $\pmb p$ and summing over ${\pmb p}$ and $\sigma$,
one finds
\begin{align}
&
\frac{\partial {\pmb G}_{\rm e}}{\partial t}=-\partial_k \Pi_{ik}^{({\rm e})} 
+ e_{\rm e} n_{\rm e} \, {\pmb E} + \frac{1}{c} \, {\pmb j}_{\rm e} \times {\pmb B},
&
\label{mome}
\end{align}
where $\Pi_{ik}^{({\rm e})}$ is given by equation~(\ref{Piik});
${\pmb G}_{\rm e}=\sum_{\pmb p \sigma} {\pmb p} \, n_{\pmb p\sigma}$
is the electron momentum density; and 
${\pmb j}_{\rm e}= \sum_{\pmb p \sigma} e_{\rm e} {\pmb v} \, n_{\pmb p \sigma}$
is the electron charge current density.
Using the Maxwell's equations, the sum of the last two terms in the right-hand sides 
of equations~(\ref{mom}) and (\ref{mome})
can be transformed, in a standard way, as
\begin{align}
&
e_{\rm p} n_{\rm p}  
\, {\pmb E} +\frac{e_{\rm p}}{c} \, n_{\rm p} \, {\pmb V}_{{\rm sp}}\times {\pmb B}
+e_{\rm e} n_{\rm e} \, {\pmb E} + \frac{1}{c} \, {\pmb j}_{\rm e} \times {\pmb B}
&
\nonumber\\
&
=\frac{1}{4\pi}\, \partial_i \left[ E_i E_k - \frac{1}{2} E^2 \delta_{ik} 
+B_i B_k -\frac{1}{2} B^2 \delta_{ik}\right] 
&
\nonumber\\
&
- \frac{1}{4 \pi c}\, \frac{\partial }{\partial t}
\left( {\pmb E}\times {\pmb B} \right).
&
\label{good}
\end{align}
Summing up equations~(\ref{mom}), (\ref{mom_n}), (\ref{mome}),
and making use of equation~(\ref{good}), 
we finally arrive at the total momentum conservation equation for our system,%
%
\footnote{In contrast to equation (\ref{momentum}), this momentum conservation equation
does not include the external force density ${\pmb f}_{\rm ext}$, applied to the vortex.
This is justified since in what follows the equation (\ref{mom_tot}) will be used at $r\gg \lambda$ 
(i.e., far from the vortex core).
} 
%
%
\begin{align}
&
\frac{\partial {\pmb G}}{\partial t}=-\partial_k \Pi_{ik}
&
\label{mom_tot}
\end{align}
with 
\begin{align}
&
\Pi_{ik}=
m_{\rm p} n_{\rm p} \, V_{{\rm sp}\, i} V_{{\rm sp} \, k}
+m_{\rm n} n_{\rm n} \, V_{{\rm sn}\, i} V_{{\rm sn} \, k}
+P_{\rm nuc} \, \delta_{ik}
&
\nonumber\\
&
\quad\,\,\,
+\Pi_{ik}^{({\rm e})}
+\Pi_{ik}^{({\rm e-m})},
&
\label{Piik2}
\end{align}
where
${\pmb G}={\pmb G}_{\rm p}+{\pmb G}_{\rm n}+{\pmb G}_{\rm e}
+{\pmb E}\times{\pmb B}/(4 \pi c)$ is the total momentum density;
$P_{\rm nuc}$ is the total neutron-proton pressure, such that  
${\pmb \nabla } P_{\rm nuc}=(n_{\rm p} \, {\pmb \nabla} \mu_{\rm p}
+n_{\rm n} \, {\pmb \nabla} \mu_{\rm n})$;
and $\Pi^{({\rm e-m})}_{ik}$ is the electromagnetic stress tensor, 
given by
\begin{align}
&
\Pi^{({\rm e-m})}_{ik}=-\frac{1}{4\pi}\,  \left[ E_i E_k - \frac{1}{2} E^2 \delta_{ik} 
+B_i B_k -\frac{1}{2} B^2 \delta_{ik}\right].
&
\label{Pie-m}
\end{align}
%

\subsection{Proof of equation~(\ref{COND2})}

Now we are able to discuss why the force on a vortex 
can be calculated from equation~(\ref{COND2}).
Below
any thermodynamic quantity $A$ is presented as
$A=A_0+\delta A$, where $A_0$ is its value in the absence of a vortex 
(note that ${\pmb \nabla}A_0=0$) 
and $\delta A$ is the vortex-related perturbation.

In the system without a vortex the electron distribution function 
is $n_{\pmb p \sigma}^{({\rm eq})}$ (see equation \ref{npeq}). 
The associated electron charge density, 
$e_{\rm e} n_{\rm e0} = \sum_{\pmb p\sigma} e_{\rm e} n_{\pmb p \sigma}^{({\rm eq})}$,
is neutralized by the background proton charge density, $e_{\rm p} n_{{\rm p}0}$.
However, when we add a vortex line to the system, 
an additional contribution $\delta n_{\pmb p\sigma}^{({\rm sc})}$ 
will arise to the electron distribution function due to electron scattering off the vortex line.
This contribution generates 
a non-zero charge and charge current densities, 
$e_{\rm e} \delta n^{({\rm sc})}$ and $e_{\rm e} \delta{\pmb j}^{({\rm sc})}$,
even far from the vortex.
Indeed, summing up equation~(\ref{NSC}) over momenta and spins, 
one obtains for $\delta n^{({\rm sc})}$
%
\footnote{
Similar expression can also be written out for $e_{\rm e} \delta {\pmb j}^{({\rm sc})}$.
However, it is easy to verify (see the text after equation \ref{cond2}) 
that the presence of this current 
does not lead to additional momentum flux through the 
boundary of a cylinder 
defined in Fig.~\ref{vortvel}
(i.e., does not affect the force ${\pmb F}_{\rm npe \rightarrow V}$).
}
%
%
\begin{align}
&
\delta n^{({\rm sc})} = \sum_{\pmb p \sigma} \delta n_{\pmb p\sigma}^{({\rm sc})}
&
\nonumber\\
&
=\alpha_1 \frac{({\pmb u}_{\rm e}-{\pmb V}_{\rm L})\cdot {\pmb e}_r}{r}
+\alpha_2 \frac{[{\pmb e}_z\times ({\pmb u}_{\rm e}-{\pmb V}_{\rm L})]\cdot {\pmb e}_r}{r},
&
\label{dnsc}
\end{align}
where ${\pmb e}_r$ is the unit vector along $r$; $\alpha_1$ and $\alpha_2$ 
are some density-dependent non-vanishing scalars, 
which can be explicitly calculated from equation (\ref{NSC}),
but are not important for the subsequent consideration.

The presence of uncompensated electron charge, $e_{\rm e} \delta n^{({\rm sc})}$,
produces an electric field ${\pmb E}$,
which affects the distribution function of incident electrons 
(so that the total distribution function is given by equation~\ref{np1})
and slightly changes the density of background protons.
All these effects should be accounted for self-consistently.
In what follows we shall work in
the coordinate system, in which vortex is at rest
(${\pmb V}_{\rm L}=0$).
We assume that
`transport' velocities of incident electrons, protons, and neutrons are small
in this coordinate system
and retain only the terms linear in the velocities in our calculations.

We start with the derivation of the expression for the induced electron distribution function 
$\delta n_{\pmb p \sigma}^{({\rm ind})}$ far from the vortex 
(at a distance $l\gg r \gg \lambda$).
At such $r$ the magnetic field and proton superfluid velocity, generated by the vortex 
in the absence of transport electron and proton currents 
are exponentially suppressed and can be neglected.
Then the kinetic equation (\ref{boltzmann}) can be written,
in the linearized form, as
\begin{align}
&
{\pmb v} \frac{ \partial \delta n_{\pmb p\sigma}^{({\rm ind})}}{\partial {\pmb r}}
+e_{\rm e}{\pmb E} \, \frac{\partial n_{\pmb p\sigma}^{({\rm eq})}}{\partial {\pmb p}}=0,
&
\label{dnpind}
\end{align}
where we 
take into account 
that the system is stationary. 
Note that, because our problem is linear,
the correction $\delta n_{\pmb p \sigma}^{\rm (sc)}$ does not
appear in this equation
(it satisfies an independent equation \ref{kinetic}).
The solution to equation (\ref{dnpind}), vanishing at $r \rightarrow \infty$, is
\begin{align}
&
\delta n_{\pmb p\sigma}^{({\rm ind})}=\frac{e_{\rm e} \phi}{v}\, 
\frac{\partial n_{\pmb p \sigma}^{({\rm eq})}}{\partial p},
&
\label{dnpind2}
\end{align}
where ${\pmb E}=-{\pmb \nabla} \phi$ and it is assumed that the electrostatic potential $\phi$ 
vanishes at $r \rightarrow \infty$.
The corresponding contribution to the electron stress tensor is
\begin{align}
&
\delta \Pi_{ik}^{({\rm e, \, ind})} = \sum_{\pmb p\sigma} \, p_i v_k \, 
\delta n_{\pmb p\sigma}^{({\rm ind})}=-e_{\rm e} n_{\rm e0}  \phi \, \delta_{ik},
&
\label{dPind}
\end{align}
while the related electron density perturbation is
\begin{align}
&
\delta n^{({\rm ind})}=\sum_{\pmb p \sigma} \delta n_{\pmb p\sigma}^{({\rm ind})} 
= \alpha_3 e_{\rm e} \phi,
&
\label{dnind}
\end{align}
where the actual form of the density-dependent parameter $\alpha_3$ is not important for us.

We turn now to calculation of the pressure perturbation, $\delta P_{\rm nuc}$, 
and the proton number density perturbation, $\delta n_{\rm p}$,
caused by the electric field ${\pmb E}$.
From the stationary equation~(\ref{sfl}) and its analogue for neutrons it follows that, 
in the linear approximation in velocities, 
\begin{align}
&{\pmb \nabla} \mu_{\rm p}=e_{\rm p} {\pmb E},&
\label{mup}\\
&{\pmb \nabla} \mu_{\rm n}=0.&
\label{mun}
\end{align}
Hence, from the definition
${\pmb \nabla } P_{\rm nuc}=n_{\rm p} \, {\pmb \nabla} \mu_{\rm p}
+n_{\rm n} \, {\pmb \nabla} \mu_{\rm n}$,
one has
\begin{align}
&
{\pmb \nabla } P_{\rm nuc} = e_{\rm p}n_{\rm p} {\pmb E},
&
\label{Pnuc}
\end{align}
and thus
\begin{align}
&
\delta P_{\rm nuc}=-e_{\rm p} n_{\rm p0} \phi,
&
\label{dPnuc}
\end{align}
where we replaced $n_{\rm p}$ with the unperturbed proton number density $n_{\rm p0}$,
which is justifiable in the linear approximation.
The perturbation $\delta n_{\rm p}$ can be found by noticing that 
$\mu_{\rm n}$ and $P_{\rm nuc}$
are the functions of $n_{\rm n}$ and $n_{\rm p}$, hence
\begin{align}
&{\pmb \nabla}\mu_{\rm n}= \frac{\partial \mu_{\rm n}}{\partial n_{\rm n}} \, {\pmb \nabla }n_{\rm n}
+\frac{\partial \mu_{\rm n}}{\partial n_{\rm p}}\, {\pmb \nabla }n_{\rm p}=
\frac{\partial \mu_{\rm n}}{\partial n_{\rm n}} \, {\pmb \nabla }\delta n_{\rm n}
+\frac{\partial \mu_{\rm n}}{\partial n_{\rm p}}\, {\pmb \nabla }\delta n_{\rm p},&
\label{mun1}\\
&{\pmb \nabla}P_{\rm nuc}= \frac{\partial P_{\rm nuc}}{\partial n_{\rm n}}\, {\pmb \nabla }n_{\rm n}
+\frac{\partial P_{\rm nuc}}{\partial n_{\rm p}} \, {\pmb \nabla }n_{\rm p}
&
\nonumber\\
&
\quad\quad\,\,\,\,=
\frac{\partial P_{\rm nuc}}{\partial n_{\rm n}}\, {\pmb \nabla }\delta n_{\rm n}
+\frac{\partial P_{\rm nuc}}{\partial n_{\rm p}} \, {\pmb \nabla }\delta n_{\rm p},&
&
\label{Pnuc1}
\end{align}
where all the partial derivatives
are taken in the unperturbed matter (in the vortex-free matter)
and we used the fact that ${\pmb \nabla} n_{\rm n, \, p}={\pmb \nabla} \delta n_{\rm n, \, p}$.
Plugging (\ref{mun1}) and (\ref{Pnuc1}) into equations (\ref{mun}) and (\ref{Pnuc}),
one derives a system of two equations for two unknown quantities, 
${\pmb \nabla}\delta n_{\rm n}$ and ${\pmb \nabla}\delta n_{\rm p}$.
The solution to this system 
allows one to relate $\delta n_{\rm p}$ with the electrostatic potential $\phi$,
\begin{align}
&
\delta n_{\rm p}=\alpha_4 e_{\rm p} \phi,
&
\label{dnp}
\end{align}
where $\alpha_4$ is a combination of partial derivatives 
from equations~(\ref{mun1}) and (\ref{Pnuc1}) (we are not interested in its exact form).
It remains to find $\phi$ using the Maxwell's equation,
\begin{align}
&
{\pmb \nabla}\cdot {\pmb E}=-\Delta \phi=4 \pi (e_{\rm e} \delta n^{({\rm sc})}+ e_{\rm e} \delta n^{({\rm ind})} + 
e_{\rm p} \delta n_{\rm p}).
&
\label{max1}
\end{align}
On the right-hand side here we see the total charge density;  
the last two terms are proportional to $\phi$ and describe reaction of the system to the perturbation 
$e_{\rm e} \delta n^{({\rm sc})}$ (the first term).
One may show that the asymptotic solution to this equation, valid at $r \rightarrow \infty$,
corresponds to vanishing right-hand side of (\ref{max1}), 
$e_{\rm e} \delta n^{({\rm sc})}+ e_{\rm e} \delta n^{({\rm ind})} + 
e_{\rm p} \delta n_{\rm p}=0$, that is 
(see equations \ref{dnind} and \ref{dnp})
\begin{align}
&
\phi = -\frac{e_{\rm e} \delta n^{({\rm sc})}}{e_{\rm e}^2 \alpha_3+e_{\rm p}^2 \alpha_4}
+O \left(\frac{1}{r^2} \right).
&
\label{}
\end{align}
As expected, $\phi \propto 1/r$ at $r \rightarrow \infty$.

Now we have everything at hand to prove equation~(\ref{COND2}).
Consider a general expression (\ref{cond1}) for the force 
${\pmb F}_{{\rm npe}\rightarrow {\rm V}}$
with $\Pi_{ik}$ given by equation (\ref{Piik2}).
Since in the absence of the vortex ${\pmb F}_{{\rm npe}\rightarrow {\rm V}}=0$,
we can rewrite (\ref{cond1}) as
\begin{align}
&
{\pmb F}_{{\rm npe}\rightarrow {\rm V}}=-\oint \delta \Pi_{ik}\, n_k \, {\rm d} S,
&
\label{cond2}
\end{align}
where $\delta \Pi_{ik}$ contains 
only vortex-related quantities.
In the linear approximation the first two velocity-dependent terms in equation~(\ref{Piik2}) 
can be omitted%
%
\footnote{
Recall that we work in the coordinate frame 
comoving with the vortex. 
In this frame the proton superfluid velocity at $r\gg \lambda$ can be presented as
${\pmb u}_{\rm e}+\delta {\pmb V}_{\rm sp}-{\pmb V}_{\rm L}$,
where ${\pmb u}_{\rm e}$ is the asymptotic proton (and electron) velocity 
far from the vortex,
and $\delta {\pmb V}_{\rm sp} =O(|{\pmb u}_{\rm e}-{\pmb V}_{\rm L}|)$
is the small perturbation induced by the charge current $\delta{\pmb j}^{({\rm sc})}$ 
of scattered electrons 
(note that the quantity $\delta {\pmb V}_{\rm sp}$
and the associated perturbation of the magnetic field  
can both be expressed through $\delta{\pmb j}^{({\rm sc})}$
using Ampere's law and equation \ref{B-u}).
Consequently, the contribution from the term 
$m_{\rm p}n_{\rm p}V_{{\rm sp}i}V_{{\rm sp}k}$ to (\ref{cond2}) 
is of the order of $O(|{\pmb u}_{\rm e}-{\pmb V}_{\rm L}|^2)$ and indeed can be neglected.
}
%
and $\delta \Pi_{ik}$ can be presented as
\begin{align}
&
\delta \Pi_{ik} \approx
\delta P_{\rm nuc} \, \delta_{ik}
+\delta \Pi_{ik}^{({\rm e, \, sc})}
+\delta \Pi_{ik}^{({\rm e, \, ind})}
&
\nonumber\\
&
\quad\quad
-\frac{1}{4\pi}\, 
\left[ E_i E_k - \frac{1}{2} E^2 \delta_{ik} 
+B_i B_k -\frac{1}{2} B^2 \delta_{ik}\right],
&
\label{deltaPiik}
\end{align}
where the electron contributions $\delta \Pi_{ik}^{({\rm e, \, sc})}$ and
$\delta \Pi_{ik}^{({\rm e, \, ind})}$ 
are given by equations~(\ref{dPi}) and (\ref{dPind}), respectively, 
and we used equation~(\ref{Pie-m}) to express 
the electro-magnetic stress tensor $\Pi_{ik}^{({\rm e-m})}$.
The first and the third terms here cancel each other out
in view of equations~(\ref{dPind}), (\ref{dPnuc})
and the quasineutrality condition for unperturbed matter, $n_{\rm e0}=n_{\rm p0}$.
Because $\phi\sim 1/r$ at large $r$, 
the terms depending on the electric field ${\pmb E}$ make a negligible 
contribution 
to ${\pmb F}_{{\rm npe}\rightarrow {\rm V}}$ and can also be ignored.
Finally, due to the symmetry of the problem, 
the magnetic field induced by the current $\delta {\pmb j}^{({\rm sc})}$ of scattered electrons, 
can only be directed along the axis $z$,
and hence also does not contribute to ${\pmb F}_{{\rm npe}\rightarrow {\rm V}}$,
which lies in the $xy$-plane.
Thus, the only non-vanishing contribution to the force comes from 
the term $\delta \Pi_{ik}^{({\rm e, \, sc})}$ (scattered electrons),
that is equation~(\ref{COND2}) is proved.

\section{Derivation of equation~(\ref{SIGMA_PAR3})   }
\label{integral}

Inserting delta-function in equation~(\ref{sigma_par2}), it can be rewritten as
\begin{align}
&
\sigma_{\|} 
= \frac{1}{2}\int_{-\infty}^{+\infty}\int_{-\infty}^{+\infty} \gamma(b)
\gamma(\tilde{b}) \,
\delta(b
-\tilde{b}) \, {\rm d} b {\rm d} \tilde{b}=
	&
\nonumber\\
&
\frac{1}{2}\int_{-\infty}^{+\infty}\int_{-\infty}^{+\infty} \gamma(b)
\gamma(\tilde{b}) \,
\frac{1}{2\pi} \int_{-\infty}^{+\infty} e^{i q (b-\tilde{b})} \, {\rm d}q
\, {\rm d} b {\rm d} \tilde{b}=
&
\nonumber\\
&
=\frac{1}{4 \pi}\, \int_{-\infty}^{+\infty}
\left|\int_{-\infty}^{+\infty} \gamma(b) \,e^{-i q b} \, {\rm d}b \right|^2 \,\, {\rm d}q 
&
\label{sigma_par5}
\end{align}
Taking into account equation~(\ref{gamma}),
(\ref{sigma_par5}) coincides with equation (\ref{SIGMA_PAR3}).

\section{Equivalence of the classical and quasiclassical expressions for $\sigma_\|$ and $\sigma_\perp$}
\label{equiv}

Let us show that expressions (\ref{sigma_par7}) and (\ref{sigma_perp7}) are equivalent to, respectively,
equations (\ref{sigma_par2}) and (\ref{sigma_perp2}) 
and thus lead to the same $\sigma_\|$ and $\sigma_\perp$.
With this aim we note that the angular momentum $\hbar l$ of a quasiclassical electron 
is related to its impact parameter $b$ by the formula $\hbar l=\hbar  k_\perp b$.
Then, introducing the coordinate $y=\sqrt{r^2-b^2}$, one rewrites equation (\ref{dl}) as
\begin{align}
&
\delta_l = - \zeta \, \frac{e_{\rm e}}{e_{\rm p}} \, b \, \int_{0}^{\infty} 
\frac{\mathcal{P}(\sqrt{b^2+y^2})}{b^2+y^2} \, {\rm d}y
&
\nonumber\\
&
\,\,\,\,
=- \zeta \, \frac{e_{\rm e}}{e_{\rm p}} \, \int_{0}^{\infty} 
\frac{\mathcal{P}(b \sqrt{1+\widetilde{y}^2})}{1+\widetilde{y}^2} \, {\rm d}\widetilde{y},
&
\label{dl2}
\end{align}
where in the second equality we changed the variable $y \rightarrow \widetilde{y}\equiv y/b$.
The derivative ${\rm d}\delta_l/{\rm d}l$ can now be calculated as
\begin{align}
&
\frac{{\rm d}\delta_l}{{\rm d}l}=\frac{1}{k_\perp} \, \frac{{\rm d}\delta_l}{{\rm d}b}
=-\frac{1}{k_\perp} \, \zeta \, \frac{e_{\rm e}}{\rm e_{\rm p}} \, 
\int_{0}^\infty \, 
\frac{{\rm d}\mathcal{P}(r)}{{\rm d}r}  \frac{{\rm d}\widetilde{y}}{\sqrt{1+\widetilde{y}^2}} 
&
\nonumber\\
&
=\frac{1}{\hbar k_\perp} \, \int_{0}^\infty \frac{e_{\rm e}}{c} \, B(\sqrt{b^2+y^2}) \, {\rm d} y,
&
\label{ddl}
\end{align}
where we make use of equation (\ref{BBB}) in the last equality.
Comparing now equations (\ref{ddl}) and (\ref{gamma}) one verifies that
$2 {\rm d}\delta_l/{\rm d}l=\gamma$ (which is an expected result, see, e.g., \citealt{ll77}) 
and hence the 
expressions (\ref{sigma_par7}) and (\ref{sigma_perp7}) coincide with, respectively,
expressions (\ref{sigma_par2}) and (\ref{sigma_perp2}).

\section{Analysis of the expression for the transverse force~${\pmb F}_\perp$}
\label{Fm}

Assume that protons form a strong type-II superconductor ($\xi \ll \lambda$).
Our aim here is 
to 
further justify 
that in this case electrons do not act {\it directly}
on the proton vortex
with the 
transverse force 
${\pmb F}_\perp$ 
(see Section \ref{disc} for a definition of ${\pmb F}_\perp$).

\begin{figure}
	\begin{center}
		\includegraphics[width=0.3\textwidth]{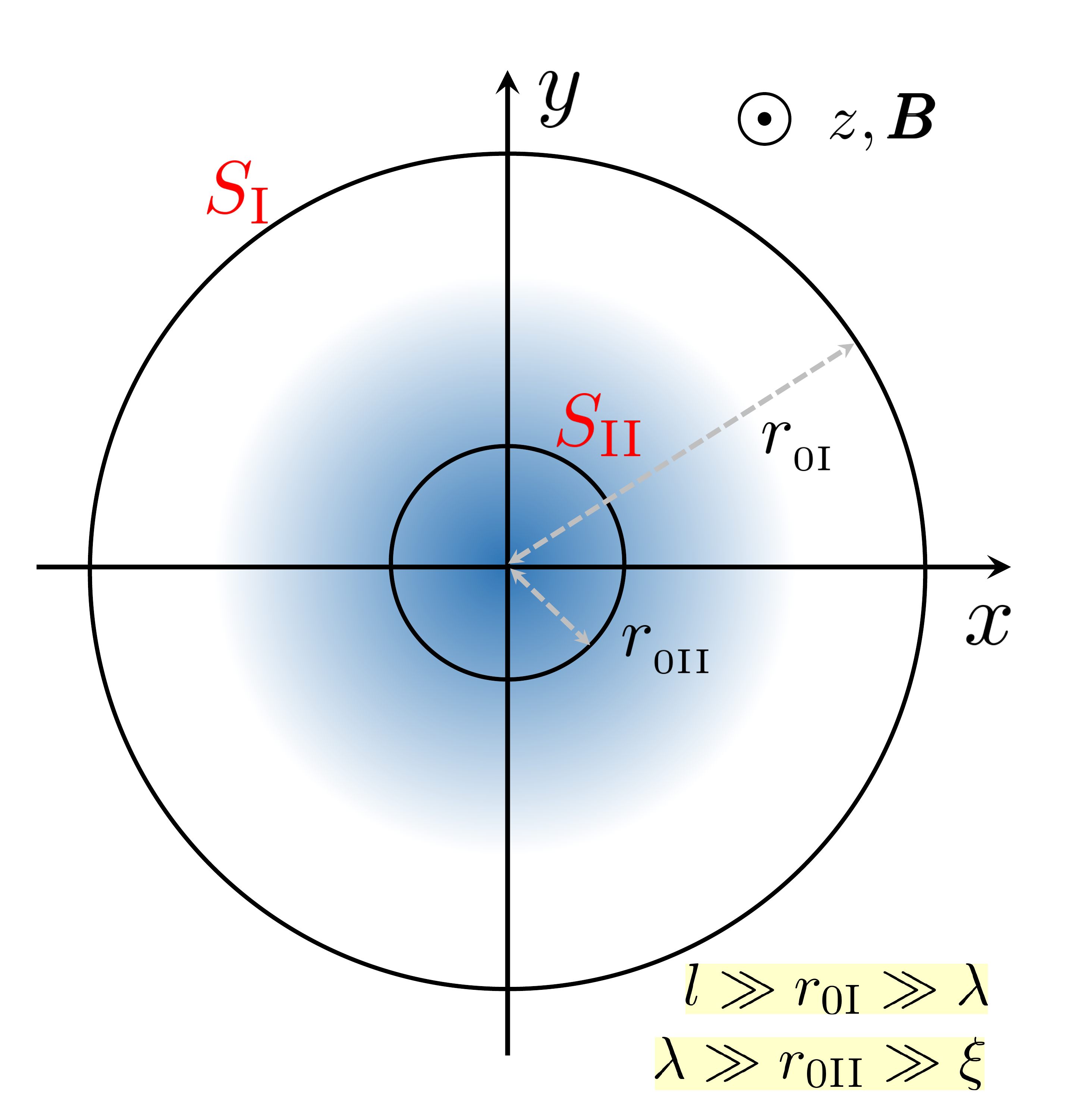}
		\caption{\label{2cyl}
			Two cylindrical surfaces, $S_{\rm I}$ and $S_{\rm II}$, 
			used to perform integration in equation (\ref{cond1})
			in order to calculate force on a vortex.
		}
	\end{center}
\end{figure}

The force on a vortex ${\pmb F}_{\rm npe\rightarrow V}$ 
is given by the general equation (\ref{cond1}).
The surface integral in this equation can be taken over any closed surface around the vortex,
which is sufficiently far from the region where the external force ${\pmb F}_{\rm ext}$ is applied.
In what follows, to take the integral, 
we consider two cylindrical surfaces of unit length, $S_{\rm I}$ and $S_{\rm II}$ (see Fig.\ \ref{2cyl}).
Let the cylindrical surface $S_{\rm I}$ be the same as in Section \ref{formulation}, 
i.e., have a radius $l \gg r_{\rm 0 \,I}\gg \lambda$;
in turn, assume that the cylindrical surface $S_{\rm II}$ 
has a radius $\lambda \gg r_{\rm 0 \, II}\gg \xi$.
Then ${\pmb F}_{\rm npe\rightarrow V}$ can be presented in two equivalent ways,
\begin{align}
&
{\pmb F}_{\rm npe \rightarrow V}=-\oint_{S_{\rm I}} \Pi_{ik} n_k {\rm d}S=
-\oint_{S_{\rm II}} \Pi_{ik} n_k {\rm d}S,
&
\label{F3}
\end{align}
where $\Pi_{ik}$ is the stress tensor for npe-matter given by equation (\ref{Piik2}). 
%
%
It consists of the contributions from the neutrons, protons, electrons, and electromagnetic field.
As shown in Section \ref{formulation}, only electrons contribute 
to the integral over the surface $S_{\rm I}$ (see equation \ref{COND2}). 
And what is the electron contribution to the integral over $S_{\rm II}$?
Using equation (\ref{mome}) (in which $\partial {\pmb G}_{\rm e}/\partial t=0$),
one can write
\begin{align}
&
-\oint_{S_{\rm II}} \Pi_{ik}^{({\rm e})} n_k {\rm d}S 
= -\int \nabla_{k}\Pi_{ik}^{({\rm e})} \, {\rm d}V
&
\nonumber\\
&
=-\int \left(e_{\rm e} n_{\rm e} {\pmb E}+\frac{1}{c}{\pmb j}_{\rm e}\times {\pmb B} \right){\rm d} V,
&
\label{F4}
\end{align}
where ${\pmb j}_{\rm e}=\sum_{{\pmb p\sigma}}e_{\rm e} \, {\pmb v} \, n_{\pmb p \sigma}$,
$n_{\rm e}=\sum_{\pmb p \sigma} n_{\pmb p \sigma}$, 
and integration is performed over the cylinder of radius $r_{\rm 0 \, II}$.
The integral (\ref{F4}) is much smaller than the corresponding electron contribution
to the momentum flux through
the surface $S_{\rm I}$, as it is demonstrated below.
Indeed, neglecting the modification of $n_{\pmb p\sigma}$, ${\pmb E}$, and ${\pmb B}$
caused by the electrons scattered off the vortex 
(i.e., using the `shifted Fermi-sphere', $n_{\pmb p \sigma}^{({\rm eq})}$,
as an electron distribution function, see equation \ref{npeq}),
one can easily perform volume integration in equation (\ref{F4}) and approximately write:
%
\footnote{Similar approximate calculation of the electron momentum flux through the 
surface $S_{\rm I}$ would give
\begin{align}
&
-\oint_{S_{\rm I}} \Pi_{ik}^{({\rm e})} n_k {\rm d}S 
\approx - \frac{e_{\rm e}}{c}  n_{\rm e} \,\Phi \, 
[({\pmb u}_{\rm e}-{\pmb V}_{\rm L})\times {\pmb e}_{z}].
&
\label{Int0}
\end{align}
This expression coincides with the transverse part ${\pmb F}_{\perp}$ of the force on a vortex 
(see Section \ref{disc} and equation \ref{Fperp}).
This means that our approximation of unperturbed quantities 
$n_{\pmb p\sigma}$, ${\pmb E}$, and ${\pmb B}$
correctly reproduces the transverse force on a vortex.
However, to calculate the dissipative longitudinal part, ${\pmb F}_\|$,
one needs to account for small deviations of 
these quantities caused by the electron scattering off the vortex magnetic field.
\label{foot}
}
%
%
\begin{align}
&
-\oint_{S_{\rm II}} \Pi_{ik}^{({\rm e})} n_k {\rm d}S 
\approx - \frac{e_{\rm e}}{c}  n_{\rm e} \,\Phi_{ S_{\rm II}}\, 
[({\pmb u}_{\rm e}-{\pmb V}_{\rm L})\times {\pmb e}_{z}],
&
\label{Int}
\end{align}
where 
\begin{align}
&
\Phi_{S_{\rm II}} =  \int_0^{r_{\rm 0\, II}} B \, 2\pi r {\rm d}r
&
\label{phiS}
\end{align}
is the magnetic flux enclosed by the cylindrical surface $S_{\rm II}$.
On the other hand, the `transverse' component of the electron momentum flux 
through the surface $S_{\rm I}$ is given by 
(see Section \ref{disc} and equation \ref{D'2} there; see also footnote \ref{foot})
\begin{align}
&
{\pmb F}_\perp =D' [{\pmb e}_z\times({\pmb u}_{\rm e}-{\pmb V}_{\rm L})]
=\frac{e_{\rm e}}{c} n_{\rm e} \Phi \, [{\pmb e}_z\times({\pmb u}_{\rm e}-{\pmb V}_{\rm L})].
&
\label{Fperp}
\end{align}
The ratio of the integrals (\ref{Int}) and (\ref{Fperp}) equals 
$\Phi_{\rm S_{\rm II}}/\Phi \sim r_{\rm 0\, II}^2/\lambda^2 \ll 1$,
i.e. the electron contribution to the momentum flux through 
the surface $S_{\rm II}$ is $r_{\rm 0\, II}^2/\lambda^2$ times smaller than through $S_{\rm I}$.
But the {\it total} momentum flux through any of these surfaces 
must be conserved (see equation \ref{F3}). 
It is clear, therefore,  
that the missing momentum flux through $S_{\rm II}$ 
should be transferred by protons. 
This 
result 
can be easily obtained
from the analysis of the proton superfluid and momentum conservation equations 
(\ref{sfl}) and (\ref{mom}) 
if we note that
the proton velocity scales as $1/r$ at distances $r\sim r_{\rm 0\, II}$, 
i.e., $\lambda \gg r \gg\xi$.
Following the same derivation as, e.g., in \cite{sonin87}, 
one then finds that
the proton contribution to the momentum flux through $S_{\rm II}$
is the ordinary Magnus force ${\pmb F}_{\rm M}$ (see equation \ref{Fmagnus}),
which coincides with the force ${\pmb F}_{\perp}$ given by equation (\ref{Fperp}) -- 
exactly what we need 
to restore momentum conservation!
We come to conclusion that the momentum flux through the surface $S_{\rm II}$
is mainly transported by protons,
while through the surface $S_{\rm I}$ -- by electrons.
In other words, 
it is the protons (not electrons),
which directly act on the vortex with the transverse force ${\pmb F}_\perp$.


\label{lastpage}

\end{document}